\documentclass[aps,prd,nofootinbib,english,10pt]{revtex4-2}
\usepackage[utf8]{inputenc}

\usepackage{graphicx,color} 
\usepackage{xcolor}
\usepackage{rotating}
\usepackage{amssymb}
\usepackage{amsfonts}
\usepackage{amsmath}
\usepackage{xspace}
\usepackage{mathtools} 
\usepackage{verbatim}
\usepackage{comment}
\usepackage{enumitem}
\usepackage{physics}
\usepackage{graphicx}
\usepackage[colorlinks]{hyperref}
\usepackage{MnSymbol}
\usepackage{graphicx}
\usepackage{accents}
\usepackage{cancel}
\usepackage{hyperref}

\begin{document}
\title{Unified field theory from Hamilton cotangent bundle geometry - The Einstein-Maxwell system} 

\author{Christian Pfeifer}
\email{christian.pfeifer@zarm.uni-bremen.de}
\affiliation{ZARM, University of Bremen, 28359 Bremen, Germany.}

\author{José Javier  Relancio}
\email{jjrelancio@ubu.es}
\affiliation{Departamento de Matemáticas y Computación, Universidad de Burgos, 09001 Burgos, Spain}
\affiliation{Centro de Astropart\'{\i}culas y F\'{\i}sica de Altas Energ\'{\i}as (CAPA),
Universidad de Zaragoza, Zaragoza 50009, Spain}

\begin{abstract}
The unification of all physical fields into one mathematical object and the derivation of all physical field equations from that object in one framework is a long-lasting endeavor in fundamental physics. We suggest a new approach to achieve this goal by encoding physical fields into the geometry of the 1-particle phase space on spacetime (the cotangent bundle) through Hamilton geometry. The fundamental field, which contains information about all physical fields in spacetime and defines the phase space geometry, is a scalar field in phase space that is interpreted as a point-particle Hamiltonian. We construct an action principle for scalar fields in phase space and derive the corresponding scalar field equation. By choosing a specific scalar field, namely the Hamiltonian describing a charged particle in curved spacetime with an electromagnetic field, we show that this phase-space scalar field equation is equivalent to the coupled Einstein-Maxwell equations in spacetime, thus providing a geometric unification of gravity and electromagnetism. We further discuss how this approach differs from previous unification attempts and its potential for describing further physical fields and their dynamics in a unified manner in terms of phase-space geometry.
\end{abstract}

\maketitle


\section{Introduction}
A fundamental unified principle, from which all equations of motion for all physical fields can be derived, is one of the most sought-after frameworks in fundamental physics. Early developments of this endeavor were performed to understand gravity and electromagnetism from a geometric perspective. Eddington~\cite{Eddington1921}, Weyl~\cite{Weyl1922}, Cartan~\cite{Cartan:1924yea} and Einstein \cite{Einstein1929} added additional geometric features, such as torsion or an independent (not metric-compatible) connection, to the geometry of a four-dimensional spacetime and attempted to interpret parts of these additional geometric structures as electromagnetic fields. The central problem with these constructions was that, when interpreted as unified field theories, they either implied a path dependence of vector lengths, led to too many fields that could not be interpreted physically, implied overdetermined field equations, or led to inconsistencies between the fields and their sources. Since then, several works have addressed this problem: using torsion~\cite{Ringermacher_1994,Kohler:1998ah}, non-metricity~\cite{Janssen:2018exh}, or other approaches~\cite{BEACH2024169661}, see~\cite{Hehl:1994ue,Vizgin1994,Goenner:2004se} for reviews on the topic.

Another famous approach to unify electromagnetism and gravity in one framework goes back to Kaluza and Klein \cite{Kaluza_1921,Klein:1926tv}, who studied five-dimensional spacetimes and attempted to obtain the Einstein-Maxwell equations from the dynamics of a spacetime metric in higher dimensions. However, the original Kaluza-Klein theory was not successful because it predicts an extra scalar field (the radion/dilaton) that has not been observed, and it imposes unrealistic charge quantization inconsistent with experiments.

From a modern perspective, the most famous framework, which further develops many of the earlier ideas to unify electromagnetism, gravity, and the entire standard model of particle physics, is String Theory. It seeks to describe all physical fields as excitations of strings, $1$-dimensional objects, moving and interacting in a higher-dimensional ambient Minkowski spacetime \cite{Polchinski_1998}. One main unsolved issue in string theory is that it usually predicts too many physical fields on spacetimes, which go beyond the field content of the standard model of particle physics \cite{Mukhi:2011zz,Marchesano:2024gul}.
 
In this article, we suggest the use of relativistic point particle Hamiltonians in the 1-particle phase space (technically on the cotangent bundle of spacetime) as fundamental objects, from which all physical fields and their dynamics on spacetime can be derived. The strategy is to determine point-particle Hamiltonians dynamically, interpreted as scalar fields on the cotangent bundle, from an action-based scalar field theory. The resulting scalar field equation, obtained through the variation of the action with respect to the Hamilton function, then decays into field equations in spacetime, which determine the physical fields in spacetime that are encoded in the point particle Hamiltonian. From a spacetime perspective, a scalar degree of freedom on the cotangent bundle can contain numerous degrees of freedom, such as a spacetime metric and vector potential.

We demonstrate the success of this strategy by determining the Einstein-Maxwell equations from the scalar field equation in the phase space. As an Ansatz for a Hamilton function $H$  we use the one of a charged particle in an electromagnetic potential $A$ on a curved spacetime with spacetime metric $G$
\begin{align}\label{eq:HamGEM}
    H = G^{\mu\nu}(x)(p_\mu - e A_\mu(x))(p_\nu - e A_\nu(x))\,.
\end{align}
The spacetime fields $G$ and $A$ can be identified from the Hamiltonian through its first and second derivatives with respect to the point particles' $4$-momenta $p$ at $p=0$. The field equations for the spacetime fields are obtained from the scalar field equation on the cotangent bundle in a similar way, namely, through the fact that each power of the point particle momenta appearing in the equation has to vanish individually.

To construct an action that is spacetime diffeomorphism invariant and defines the scalar field theory on the cotangent bundle, we employ a geometric method called Hamiltonian geometry \cite{miron2001geometry,Miron_HamiltonGeometry_2012}\footnote{Hamilton geometry is a dual formulation of Lagrange geometry, and generalises Finsler geometry \cite{Rund2012,Bucataru,BCS,Hohmann:2021zbt}, as it does not require the Hamiltonians to be homogeneous in the particles' momenta.}. This mathematical framework allows us to extend the geometric principle of general relativity, where gravity is encoded into the pseudo-Riemannian geometry of spacetime determined by a spacetime metric, to Hamilton functions that determine the geometry of the cotangent bundle. Thus, if all physical field content on spacetime is encoded in Hamilton functions, the Hamilton geometry of the cotangent bundle encodes all physical interactions.

We begin this article by introducing the basics of Hamilton geometry in Section \ref{sec:HamGeom}, where we demonstrate how to write down action integrals for the geometry defining the Hamilton function in general. In Section \ref{sec:dhf} we specify the action and derive the scalar field equation that provides dynamics to the Hamilton function. We then evaluate the field equation in Section \ref{sec:Einst-Max} for the Hamiltonian of a relativistic point particle in an electromagnetic potential in curved spacetime. We explicitly show that the general scalar field equation of the Hamiltonian decays into the usual Einstein-Maxwell system for the metric and vector potential on spacetime. Finally, we conclude the article in Section \ref{sec:discussion}.

The notational conventions of this article are that indices $\nu, \mu, ..$ run from $0$ to $3$. Moreover, we use the (anti)-symmetrization brackets with exception signs $|\ |$ as follows 
\begin{align}
    X_{(\mu_1...|\rho|...\mu_n)}=\frac{1}{n!} \textrm{ (total symmetrization over all indices $\mu_1...\mu_n$)}
\end{align}
 and 
\begin{align}
     X_{[\mu_1...|\rho|...\mu_n]}=\frac{1}{n!} \textrm{(total anti-symmetrization over all indices $\mu_1...\mu_n$)}.
\end{align}

\section{The basics of Hamiltonian geometry}\label{sec:HamGeom}
The Hamilton geometry framework is the foundation for dynamically determining point-particle Hamilton functions. We briefly introduce and recall the required concepts. The main references for the development of the framework are \cite{miron2001geometry,Miron_HamiltonGeometry_2012}, where all details about the objects introduced in this section can be found. In physics, the framework has already been successfully applied to study curved spacetimes with curved momentum spaces that describe aspects of quantum gravity phenomenology \cite{Barcaroli:2015xda,Pfeifer:2021tas,Relancio:2022mia}.

Before introducing the geometric objects of Hamiltonian geometry in Section \ref{ssec:geomObj} and understanding how to write down action integrals in Section \ref{ssec:integration}, we define the cotangent bundle as a manifold in itself. The entire construction of the Hamiltonian geometry works in any dimension; however, because we will be working on a physical space-time manifold, we set the dimension of the base manifold to four.

The \emph{cotangent bundle} $T^*M$ of a $4$-dimensional spacetime manifold $M$ is the union of all cotangent spaces 
\begin{align}
    T^*M = \bigcup_{p\in M} T_{p}M\,,
\end{align}
and itself is an $8$-dimensional manifold. In addition, it is  a vector bundle that possesses a natural projection $\pi: T^*M \to M$ whose fibers at each point $p\in M$ are given by $\mathbb{R}^4\sim T^*_pM$.

We equip the cotangent bundle with the so-called \emph{manifold-induced coordinates} as follows. Consider $P\in T^*M$, then $P$ is an element of the cotangent space $T^*_{\pi(P)}M$. Let $(U,x)$ be a coordinate chart around $q = \pi(P) \in M$, then we can express $P$ in the canonical coordinate basis as $P = p_\mu dx^\mu$. The manifold induced coordinates on $\pi^{-1}(U)\subset T^*M$ are now given by associating to $P$ the $8$-numbers
\begin{align}
    P \sim (x^\mu,p_\nu)\,,
\end{align}
where $x^\mu$ are the coordinates of $q=\pi(P)\in U$ and $p_\nu$ are the coordinate components of the $1$-form $P$ at $q$ expressed in the coordinate basis of $T^*_qM$. In these manifold induced coordinates, the coordinate bases of the tangent $T_PT^*M$ and cotangent $T^*_PT^*M$ spaces of the cotangent bundle are denoted by
\begin{align}\label{eq:coordbas}
    T_PT^*M = \textrm{span}\left\{\partial_\mu = \frac{\partial}{\partial x^\mu}, \bar\partial^\nu =  \frac{\partial}{\partial p_\nu}\right\}\,,\quad
    T^*_PT^*M = \textrm{span}\left\{dx^\mu, dp_\nu \right\}\,.    
\end{align}
Under coordinate changes on the base manifold $x\to\tilde x(x)$ the manifold induced coordinates change to
\begin{align}
    (x^\mu, p_\nu) \to (\tilde x^\mu(x), \tilde p_\nu(x,p)) = (\tilde x^\mu(x), \tilde \partial_\nu x^\rho  p_\rho)\,,
\end{align}
where $\tilde \partial_\nu x^\rho$ is the inverse of $ \partial_\rho \tilde x^\nu(x)$. The transformation behavior of the coordinate bases \eqref{eq:coordbas} follows immediately. We like to highlight that, remarkably, even though being bases fields on the cotangent bundle, the objects $\bar\partial^\mu$ and $dx^\mu$ transform formaly as if they where tensor fields on the base manifold
\begin{align}
    dx^\nu \to d \tilde x^\nu = \partial_\mu \tilde x^\nu d x^\mu,\quad \bar\partial^\nu\to \tilde{\bar\partial}^\nu = {\partial}_\mu \tilde x^\nu \bar\partial^\mu \,.
\end{align}
The components of tensor fields on the cotangent bundle that share this property are called \emph{distinguished} or short \emph{d}-tensor fields. 

Another canonical structure on $T^*M$ is the \emph{Poisson bracket}. Let $f$ and $h$ be real functions on $T^*M$. Then, the Poisson bracket between $f$ and $h$ is defined as
\begin{align}\label{eq:poisson}
    \{f, h\} = \bar\partial^\lambda f \partial_\lambda h - \partial_\lambda f \bar \partial^\lambda h\,.
\end{align}

Having set the stage, we can define the geometric objects in Hamilton geometry, which will serve as building blocks for an action that determines the Hamilton function dynamically in Section \ref{sec:dhf}.

\subsection{Geometric objects}\label{ssec:geomObj}
The fundamental ingredient from which the geometry of the cotangent bundle is derived in Hamilton geometry is a Hamilton function, that is, a scalar field on the cotangent bundle. The construction works similarly to pseudo-Riemannian geometry, in which the geometry of a manifold is derived from a metric tensor in spacetime.

More precisely: Let $H$ be a real function on $T^*M$
\begin{align}
    H: T^*M&\to \mathbb{R}\\
    (x,p) &\mapsto H(x,p)\,.
\end{align}
and define the components of the so-called \emph{Hamiltonian metric} as
\begin{align}
    g^{\mu\nu}(x,p) = \frac{1}{2}\bar\partial^\mu \bar\partial^\nu H\,,
    \label{eq:metric_H}
\end{align}
Then, we call $(M,H)$ a general \emph{Hamiltonian manifold}, or \emph{general Hamiltonian space}, if the Hamiltonian metric admits an inverse $g_{\mu\nu}(x,p)$
\begin{align}
    g^{\mu\nu}g_{\mu\sigma} = \delta^\nu_\sigma\,,
\end{align}
nearly everywhere on $T^*M$. Demanding the existence of this inverse everywhere on $T^*M$ is too restrictive to cover all interesting examples that might occur, for example, in ray optics in crystals \cite{Perlick2000RayOptics}.

The geometric objects of a Hamiltonian manifold $(M,H)$ are derived as follows~\cite{miron2001geometry}:
\begin{itemize}
    \item A canonical connection structure on $T^*M$, which generalizes the Levi-Civita connection from pseudo-Riemannian geometry, which enables us to do covariant differentiation can be introduced through the \emph{non-linear connection coefficients}
    \begin{align}\label{eq:NonLin}
        N_{\mu\nu} = \frac{1}{4}\left( \{g_{\mu\nu},H \}  - g_{\mu\rho} \partial_\nu \bar \partial^\rho H - g_{\nu\rho} \partial_\mu \bar \partial^\rho H\right)\,,
    \end{align}
    where $\{\ .\ ,\ .\ \}$ is the Poisson bracket introduced in \eqref{eq:poisson}. 
    
    \item The \emph{horizontal derivative} operators are given by
    \begin{align}
        \delta_\mu = \partial_\mu + N_{\mu\nu}\bar\partial^\nu \,.
    \end{align}
    They have the property that they are $d$-vector fields on $T^*M$, that is, their transformation behavior under manifold-induced coordinate changes mimics the behavior of basis vector fields on the base manifold. Moreover, they induce a split of the tangent spaces to the cotangent bundle into horizontal and vertical parts
    \begin{align}
        T_{(x,p)}T^*M = \text{H}_{(x,p)}T^*M \oplus \text{V}_{(x,p)}T^*M\,,
    \end{align}
    with
    \begin{align}
        \text{H}_{(x,p)}T^*M = \textrm{span}\left\{ \delta_a\right\}\,,\quad \text{V}_{(x,p)}T^*M = \textrm{span}\left\{ \bar\partial^a\right\}\,.
    \end{align}
    The horizontal part can be interpreted as the cotangent bundle representation of the tangent space to the base manifold $M$, whereas the vertical part is the tangent space to the cotangent space $T^*_xM$. 
    
    \item The \emph{non-linear curvature} $R_{\rho\mu\nu}$ measures the integrability of the distribution of horizontal tangent spaces, and is given by
    \begin{align}
        R_{\rho\mu\nu} \bar\partial^\rho= [\delta_\mu,\delta_\nu] = (\delta_\mu N_{\nu\rho} - \delta_\nu N_{\mu\rho})\bar\partial^\rho\,.
        \label{eq:NonLinCurv}
    \end{align}
\end{itemize}

The canonical nonlinear connection coefficients are unique connection coefficients that ensure the constancy of the Hamiltonian functions along the horizontal directions for Hamiltonians that are homogeneous with respect to the momenta. Meaning, that Hamilton functions $H$ satisfying 
\begin{align}
    H(x,\lambda p) = \lambda^n H(x,p)\ \forall \lambda>0\,,
\end{align}
it follows that
\begin{align}
    \delta_\mu H = 0\,.
\end{align}
For such Hamiltonians, Hamilton geometry is the dual (cotangent bundle) of Finsler geometry \cite{Bucataru,BCS,Hohmann:2021zbt}, which is usually formulated on the tangent bundle.

In addition to the horizontal derivative operators, we define two covariant derivatives for tensor fields on the cotangent bundle, which allow us to write the field equations that we will obtain for the Hamilton function in Section \ref{sec:dhf} in a covariant manner.
\begin{itemize}
    \item The \emph{canonical metric compatible covariant derivative} is defined through its action on the horizontal and vertical basis
    \begin{align}\label{eq:cartanCD}
        \nabla_\mu\delta_{\nu} &= H^\sigma{}_{\mu\nu} \delta_\sigma\,,\quad  &&\nabla_{\bar\partial^\mu}\delta_{\nu} = C_\nu{}^{\mu\sigma}{}\delta_\sigma\,,\nonumber\\
        \nabla_\mu\bar\partial^{\nu} &= - H^\nu{}_{\mu\sigma} \bar\partial^\sigma\,,  &&\nabla_{\bar\partial^\mu}\bar\partial^{\nu} = - C_\sigma{}^{\mu\nu} \bar\partial^\sigma\,,\\
        H^\sigma{}_{\mu\nu} &= \frac{1}{2}g^{\sigma\lambda} (\delta_\mu g_{\lambda\nu} + \delta_\nu g_{\lambda\mu} - \delta_\lambda g_{\mu\nu} )\,,
        &&C_\sigma{}^{\mu\nu} =- \frac{1}{2}g_{\sigma\lambda} (\bar\partial^\mu g^{\lambda\nu} + \bar\partial^\nu g^{\lambda\mu} - \bar\partial^\lambda g^{\mu\nu} ) = - \frac{1}{2}g_{\sigma\lambda} \bar\partial^\mu g^{\lambda\nu}\,.\nonumber
    \end{align}
    They have the properties that
    \begin{align}
        \nabla_\mu g{}_{\rho\sigma} = 0,\quad 
        \nabla_{\bar\partial^\mu} g{}_{\rho\sigma} = 0,\quad 
        \nabla_\mu p_\nu = N_{\mu\nu} - H^\sigma{}_{\mu\nu}p_\sigma\,.
    \end{align}

    The horizontal and vertical covariant derivatives of tensor fields are given by
    \begin{equation}
    \begin{split}
           \nabla_\rho T^{\mu_1\dots\mu_n}_{\nu_1\dots\nu_n}&= \delta_\rho T^{\mu_1\dots\mu_n}_{\nu_1\dots\nu_n}+T^{\lambda\mu_2\dots\mu_n}_{\nu_1\dots\nu_n}{H^{\mu_1}}_{\lambda \rho}+\dots +T^{\mu_1\dots\lambda}_{\nu_1\dots\nu_n}{H^{\mu_n}}_{\lambda \rho}-T^{\mu_1\dots\mu_n}_{\lambda\nu_2\dots\nu_n}{H^{\lambda }}_{\nu_1 \rho}-\dots T^{\mu_1\dots\mu_n}_{\nu_1\dots\lambda}{H^{\lambda }}_{\nu_n \rho} \,,\\
         \nabla_{\bar\partial^\rho} T^{\mu_1\dots\mu_n}_{\nu_1\dots\nu_n}&= \bar\partial^\rho T^{\mu_1\dots\mu_n}_{\nu_1\dots\nu_n}+T^{\lambda\mu_2\dots\mu_n}_{\nu_1\dots\nu_n}{C_{\lambda}}^{\mu_1 \rho}+\dots +T^{\mu_1\dots\lambda}_{\nu_1\dots\nu_n}{C_{\lambda}}^{\mu_n \rho}-T^{\mu_1\dots\mu_n}_{\lambda\nu_2\dots\nu_n}{C_{\nu_1}}^{\lambda \rho }-\dots T^{\mu_1\dots\mu_n}_{\nu_1\dots\lambda}{C_{\nu_n}}^{\lambda \rho }\,,
    \end{split}
    \end{equation}
respectively.
    \item The \emph{Berwald covariant derivative}:
    \begin{align}
        \tilde \nabla_\mu\delta_{\nu} &= \bar\partial^\sigma N_{\mu\nu} \delta_\sigma\,,\quad 
        &&\tilde \nabla_{\bar\partial^\mu}\delta_{\nu} = 0\,,\\
        \tilde \nabla_\mu\bar\partial^{\nu} &= - \bar\partial^\nu N_{\mu\sigma} \bar\partial^\sigma\,, 
        &&\tilde \nabla_{\bar\partial^\mu}\bar\partial^{\nu} =  0\,.
    \end{align}
    It has the properties that
    \begin{align}
        \tilde \nabla_\mu g{}_{\rho\sigma} &\neq 0,\quad 
        \tilde \nabla_\mu p_\nu = N_{\mu\nu} - p_\sigma\bar\partial^\sigma N_{\mu\nu} \,.
    \end{align}
    \item The \emph{Landsberg tensor} characterizes the difference between the covariant derivatives
    \begin{align}\label{eq:Lands}
        S^\mu{}_{\nu\rho} = H^\mu{}_{\nu\rho} - \bar\partial^\mu N_{\nu\rho}\,.
    \end{align}
    \item Each of the covariant derivatives induces a 4-index tensor, which resembles the Riemann curvature tensor. For the purpose of this paper we do not use the full curvature tensor of the covariant derivatives, as introduced in~\cite{miron2001geometry}, but a reduction of those which was found more suitable in~\cite{Relancio:2020rys}, to find some consistent Einstein’s equations compatible with a momentum independent energy momentum tensor,
    \begin{align}\label{eq:CartanCurv}
     \mathcal{R}^\mu{}_{\nu\rho\sigma}(x,p) = \delta_\rho H^\mu{}_{\nu\sigma} - \delta_\sigma H^\mu{}_{\nu\rho} + {H^\mu}_{\lambda\rho} {H^\lambda}_{\nu \sigma} - {H^\mu}_{\lambda\sigma} {H^\lambda}_{\nu \rho}
        \,,\quad \mathcal{R} = {  \mathcal{R}^\mu}_{\nu\mu\sigma}g^{\sigma\nu}=   \mathcal{R}_{\nu\sigma}g^{\sigma\nu}\,.
    \end{align}
\end{itemize}
All the objects introduced have the property that they are $d$-tensor fields. They exhibit a well-defined covariant transformation behavior under manifold-induced coordinate changes. Therefore, they are the building blocks of the Lagrange function, which defines the action to determine the dynamics of the Hamilton functions.

Before discussing the action integral, we would like to highlight that for a Hamiltonian that is quadratic in the momenta $p$ 
\begin{align}\label{eq:HamG}
    H_G = G^{\mu\nu}(x)p_\nu p_\mu\,,
\end{align}
all the above objects reduce to that of a pseudo-Riemannian manifold with metric $G$, that is, in this case, the geometry of $(M,H_G)$ is equivalent to the geometry of $(M,G)$:
\begin{align}
    g^{\mu\nu}(x,p) = G^{\mu\nu}(x)\,,\quad 
    N_{\mu\nu} = \Gamma^\sigma{}_{\mu\nu}[G]p_\sigma\,,\quad
    R_{\mu\nu\rho}(x,p) = R^\sigma{}_{\mu\nu\rho}[G] p_\sigma\,,\quad
    \mathcal{R}^\sigma{}_{\mu\nu\rho} = R^\sigma{}_{\mu\nu\rho}[G]\,,
\end{align}
where $\Gamma^\sigma{}_{\mu\nu}[G]$ are the Christoffel symbols that define the Levi-Civita connection of $G$ and $R^\sigma{}_{\mu\nu\rho}[G]$ is the corresponding Riemannian curvature tensor. Moreover, the vertical connection coefficients and the Landsberg tensor vanish $C_{\sigma}{}^{\mu\nu} = 0,\  S^\mu{}_{\nu\rho}=0$.

\subsection{Action integrals of the cotangent bundle}\label{ssec:integration}
To write an action for the Hamiltonian functions, we need to integrate over the cotangent bundle. Compared to the construction of action integrals on the tangent bundle, which are for example employed for field theories on Finsler spacetimes \cite{Hohmann:2021zbt,Hohmann:2018rpp}, this turns out to be surprisingly simple, due to the existence of a canonical volume form $\Sigma$ on $T^*M$, that is invariant under manifold induced coordinate transformations
\begin{align}
    \Sigma = dx^0\wedge ... \wedge dx^3 \wedge dp_0 \wedge ... \wedge dp_3 =: d^4x d^4p\,.
\end{align}

Let $L$ be a scalar function on $T^*M$, which is constructed from $d$-tensor field components built from a Hamiltonian function $H$ and its derivatives, as introduced in Section \ref{ssec:geomObj}. Then, 
\begin{align}
    S[H] = \int_{T^*M} \Sigma\ L(H, \partial H, \bar \partial H, \partial \partial H, \bar \partial\partial H, \bar \partial\bar \partial H, ....)\,,
\end{align}
is an action integral that is invariant under manifold-induced coordinate transformations, from which the field equations for $H$ can be derived using the calculus of variations. It defines a scalar field theory on $T^*M$.

From the geometric objects introduced in Section \ref{ssec:geomObj}, numerous tangent bundle scalars can be constructed to build Lagrange functions that are candidates for defining the dynamics of a Hamiltonian. In the next section we identify one possible choice here, which leads to the Einstein-Maxwell equations in case the Hamiltonian is chosen as
\begin{align}\label{eq:HEdynCST}
    H = G^{\mu\nu}(x)(p_\mu - e A_\mu(x))(p_\nu - e A_\nu(x))\,,
\end{align}
as will be proven in Section \ref{sec:Einst-Max}.

A general study of the properties of the different scalar field theories on the cotangent bundle that can be constructed and their implications for unified field theories is left for future studies.

\section{Dynamical Hamilton functions - action and field equations}
\label{sec:dhf}
To provide dynamics to a Hamiltonian in terms of a scalar field theory on the cotangent bundle, we must fix a Lagrange function that defines an action. The geometric objects constructed in Section \ref{ssec:geomObj} serve as building blocks to guarantee spacetime diffeomorphism invariance of the action. The field equations will the be obtained using variational calculus. In Section \ref{sec:Einst-Max} we demonstrate that for a specific choice of Hamiltonian, these field equations on the cotangent bundle are equivalent to the Einstein-Maxwell equations.

\subsection{The Lagrangian and the action}

To identify suitable candidates for a Lagrange function $L$ we observe that, in natural units, the entire action is of mass dimension $0$. The volume form does not carry any mass-dimension, as it cancels between the products of $dx$ and $dp$ 
\begin{align}
   [dx^\mu]  = [\bar\partial^\mu] = [m^{-1}]\,, \quad [\partial_\mu] = [dp_\mu] = [m]\,.
\end{align}
Hence, the Lagrange function $L$ itself must also have a mass dimension of $0$. Fixing the dimension of the Hamiltonian to be
\begin{align}
    [H] = [m^2]\,,
\end{align}
the derived objects have the mass dimensions
\begin{align}
    [g^{\mu\nu}] = [m^0], \quad 
    [\mathcal{R}^\mu{}_{\nu\rho\sigma}]=[  \mathcal{R}_{\mu\nu}]=[\mathcal{R}] = [m^2]\,, \quad
    [R_{\mu\nu\rho}] = [m^3]\,,\quad
    [R] = [m^4]\,.
\end{align}
To ensure that for Hamilton functions $H=g^{\mu\nu}(x)p_\mu p_\nu$ we obtain the Einstein vacuum equations from the scalar field dynamics on the cotangent bundle, the action defining the Lagrangian should include a term proportional to $\mathcal{R}$. Moreover, to ensure the inclusion of the field strength tensor of the electromagnetic field $F_{\mu\nu} = \partial_{[\mu} A_{\nu]}$ in the case where a Hamiltonian of the type \eqref{eq:HEdynCST} is chosen, terms that include $\delta_\mu H$ are needed.

A minimal, dimensionless action, which, as we will see in Section \ref{sec:Einst-Max}, leads to the Einstein-Maxwell equations, contains 3-terms
\begin{align}\label{eq:act}
    S[H] = \int_{T^*M} d^4x d^4p\ \left( \alpha_1 \frac{\mathcal{R}}{H}  + \kappa \left(\alpha_2 \frac{1 }{H^2} g^{\mu\nu}\delta_\mu H \delta_\nu H  +  \alpha_3 \frac{1 }{H} (\bar\partial^\rho \delta_\mu H )(\bar\partial^\mu\delta_\rho H ) \right) \right)\,,
\end{align}
where $\kappa$ is proportional to the gravitational constant (which is of mass-dimension $[\kappa]=[m^{-2}]$), and the coupling constants $\alpha_1, \alpha_2$ and $\alpha_3$ are used as freedom to adjust the relative weight terms in the field equations.

\subsection{Variation and the field equations}

To obtain the field equations through systematic variation, we split this action into three integrals:
\begin{align}
    I_1[H] &= \int d^4x d^4p \ \frac{\mathcal{R}}{H}\,,\\
    I_2[H] &= \int d^4x d^4p \ \frac{1}{H^2} g^{\mu\nu}\delta_\mu H \delta_\nu H\,,\\
    I_3[H] &= \int d^4x d^4p \ \frac{1}{H} (\bar\partial^\rho \delta_\mu H )(\bar\partial^\mu\delta_\rho H )\,.
\end{align}
Moreover, in the first step, we isolate the variation of the nonlinear connection coefficients for all integrals $I_i$, before we continue the variation of the connection coefficients separately in the second step. As usual, the boundary terms that emerge from the integration by parts are neglected in the analysis.
\begin{itemize}
    \item The variation of $I_1$ with respect to the Hamilton function can be written as follows 
    \begin{align}
        \delta I_1 =  \int d^4x d^4p\ \left[  \left( \frac{1}{2} \bar\partial^\mu \bar\partial^\nu (\mathcal{R}_{\mu\nu}/H + Q^\rho{}_\rho g_{\mu\nu} - Q_{\mu\nu}) - \mathcal{R}/H^2  \right) \delta H + \delta N_{\nu\sigma} X^{\nu\sigma}_1 \right]\,,
    \end{align}
    where 
    \begin{align}\label{eq:X1}
        X_1^{\nu\sigma} = 2\left(\frac{S^\sigma{}_{\sigma\rho}}{H}+\frac{\delta_\rho H}{H^2}\right) \left(C^{\rho\nu\sigma} - g^{\rho(\nu}C_{\mu}{}^{\sigma)\mu}\right) +\frac{1}{H}(g^{\mu\lambda}\bar\partial^{(\nu} H^{\sigma)}{}_{\mu\lambda}  - g^{\mu(\sigma}  \bar\partial^{\nu)} H^\rho{}_{\mu\rho}) \,,
    \end{align}
    with $C$ being the vertical connection coefficient defined in \eqref{eq:cartanCD},
    \begin{equation}\label{eq:Qtensor}
        Q_{\lambda\nu}=\left(-\nabla_\nu+{S^\rho}_{\rho\nu} \right)\left(\frac{1}{H^2}\delta_\lambda H+\frac{1}{H}{S^\sigma}_{\sigma\lambda}\right)\,,
    \end{equation}
    and $S$ is the Landsberg tensor, as defined in \eqref{eq:Lands}.
    
    \item The variation of $I_2$ yields
    \begin{align}
        \delta I_2 = \int   d^4x d^4p\ \left[ \left( \frac{1}{2}\bar\partial^\mu\bar\partial^\nu \left(\frac{\delta_\mu H \delta_\nu H}{H^2}\right) - 2 \frac{g^{\mu\nu}\delta_\mu H \delta_\nu H}{H^3} - 2 \tilde \nabla_\mu \left( \frac{g^{\mu\nu}\delta_\nu H}{H^2}\right) \right) \delta H + \delta N_{\nu\sigma} X_2^{\nu\sigma} \right]  \,,
    \end{align}
    with
    \begin{align}\label{eq:X2}
        X_2^{\nu\sigma} = 2 \frac{g^{\mu(\nu}\bar\partial^{\sigma)} H \delta_\mu H}{H^2}\,.
    \end{align}

    \item And the variation of $I_3$ becomes
    \begin{align}
        \delta I_3 = \int   d^4x d^4p\ \left[ \left( -\frac{1}{H^2}(\bar\partial^\rho \delta_\mu H )(\bar\partial^\mu\delta_\rho H ) 
        + 2 \tilde\nabla_\mu \bar\partial^\rho\left(\frac{\bar\partial^\mu \delta_\rho H}{H}\right)  
        \right) \delta H + \delta N_{\nu\sigma} X_3^{\nu\sigma} \right] \,,
    \end{align}
    with
    \begin{align}\label{eq:X3}
        X_3^{\nu\sigma} = - 2  (\bar\partial^{(\sigma} H) \bar\partial^{|\rho|}\left(\frac{\bar\partial^{\nu)}\delta_\rho H}{H}\right) \,.
    \end{align}
 \end{itemize}
Summarizing the total variation of the action with respect to the Hamiltonian $H$ this far, we have that
\begin{align}
    \delta S = \int   d^4x d^4p\ &\Bigg[ 
    \alpha_1\left( \tfrac{1}{2} \bar\partial^\mu \bar\partial^\nu \left(\tfrac{\mathcal{R}_{\mu\nu}}{H} + Q^\rho{}_\rho g_{\mu\nu} - Q_{\mu\nu}\right) - \tfrac{\mathcal{R}}{H^2}  \right) \nonumber\\
    &+ \kappa \Bigg\{ \alpha_2 \left( \tfrac{1}{2}\bar\partial^\mu\bar\partial^\nu \left(\tfrac{\delta_\mu H \delta_\nu H}{H^2}\right) - 2 \tfrac{g^{\mu\nu}\delta_\mu H \delta_\nu H}{H^3} - 2 \tilde \nabla_\mu \left( \tfrac{g^{\mu\nu}\delta_\nu H}{H^2}\right) \right)\nonumber\\
    &+ \alpha_3 \left( -\tfrac{1}{H^2}(\bar\partial^\rho \delta_\mu H )(\bar\partial^\mu\delta_\rho H ) 
    + 2 \tilde\nabla_\mu \bar\partial^\rho\left(\tfrac{\bar\partial^\mu \delta_\rho H}{H}\right)  
    \right) \Bigg\}
    \Bigg]
    \delta H + X^{\mu\nu}\delta N_{\mu\nu}
\end{align}
where we abbreviated $X^{\mu\nu} = \alpha_1 X^{\mu\nu}_1+\kappa (\alpha_2 X^{\mu\nu}_2+ \alpha_3 X^{\mu\nu}_3)$.

In a second step, we now display the variation of the non-linear connection coefficients with respect to the Hamiltonian, which is a lengthy but conceptually clear calculation that leads to
\begin{align}\label{eq:defX}
    \frac{1}{4}\int d^4x d^4p\  X \delta H &:=\int   d^4x d^4p\ \delta N_{\nu\sigma} X^{\nu\sigma}\nonumber\\
    &= \frac{1}{4}\int d^4x d^4p 
    \bigg[
    -\tfrac{1}{2}\bar\partial^\sigma \bar\partial^\lambda\left(  X^{(\mu\nu)}\delta_\alpha H \bar\partial^\alpha (g_{\mu\sigma}g_{\nu\lambda})   -  \bar\partial^\alpha (X_{(\lambda\sigma)}\delta_\alpha H  )\right)\nonumber \\
    &-\tilde\nabla_\alpha \left(X^{(\mu\nu)}\bar\partial^\alpha g_{\mu\nu}\right)  - \tilde\nabla_{\mu}\bar\partial^\rho(2X^{(\mu\nu)}g_{\nu\rho})
    + \bar\partial^\lambda(2X^{(\mu\nu)}g_{\nu\rho}{S^\rho}_{\mu\lambda})\nonumber\\
    &-\tfrac{1}{2}\bar\partial^\lambda \bar\partial^\sigma \left(\tilde\nabla_\alpha(X_{(\lambda\sigma)}\bar\partial^\alpha H) -  S^\omega{}_{\alpha\lambda} X_{\omega\sigma}\bar\partial^\alpha H  -  S^\omega{}_{\alpha\sigma} X_{\omega\lambda}\bar\partial^\alpha H \right)\nonumber\\
    &+\bar\partial^\alpha \bar\partial^\beta \left(X^{(\mu\nu)}g_{\alpha\mu} g_{\beta\rho} \nabla_\nu \bar \partial^\rho H \right)
    \bigg]\delta H\,.
\end{align}
The scalar $X$ can be easily identified from the last equation. 

Writing the variations in this way we find the \emph{field equations for the Hamiltonian $H$ on the cotangent bundle} to be
\begin{align}\label{eq:HamEOM}
    &\alpha_1\left( \tfrac{1}{2} \bar\partial^\mu \bar\partial^\nu \left(\tfrac{\mathcal{R}_{\mu\nu}}{H} + Q^\rho{}_\rho g_{\mu\nu} - Q_{\mu\nu}\right) - \tfrac{\mathcal{R}}{H^2}  \right)\nonumber\\
    &+ \kappa \Bigg\{ \alpha_2 \left( \tfrac{1}{2}\bar\partial^\mu\bar\partial^\nu \left(\tfrac{\delta_\mu H \delta_\nu H}{H^2}\right) - 2 \tfrac{g^{\mu\nu}\delta_\mu H \delta_\nu H}{H^3} - 2 \tilde \nabla_\mu \left( \tfrac{g^{\mu\nu}\delta_\nu H}{H^2}\right) \right)\nonumber\\
    &+ \alpha_3 \left( -\tfrac{1}{H^2}(\bar\partial^\rho \delta_\mu H )(\bar\partial^\mu\delta_\rho H ) 
    + 2 \tilde\nabla_\mu \bar\partial^\rho\left(\tfrac{\bar\partial^\mu \delta_\rho H}{H}\right)  
    \right) \Bigg\} + \frac{1}{4}X = 0\,.
\end{align}

We will see next that, for a suitable choice of the Hamiltonian, these field equations imply the Einstein-Maxwell equations.

\section{The Einstein-Maxwell system from the field equation that determines Hamiltonians dynamically}\label{sec:Einst-Max}
We now study the field equation \eqref{eq:HamEOM} for specific classes of Hamiltonians. First, we confirm that when we choose the Hamiltonian of a free-falling particle in general relativity, $H=H_G$ \eqref{eq:HamG}, as an Ansatz, we obtain the Einstein-vacuum equations. Second, we use the Hamiltonian for a charged particle in an electromagnetic field \eqref{eq:HamGEM} to derive the Einstein-Maxwell equations from Eq. \eqref{eq:HamEOM}.

For both examples in this section, there exists a space-time metric $G^{\mu\nu}(x)$. Indices that are not in their natural position are raised or lowered using this $p$-independent metric.

\subsection{Recovering the Einstein equations}\label{ssec:GR}
As a first sanity check of the field equation \eqref{eq:HamEOM}, we consider the Hamiltonian of a free particle in curved spacetime, as given by \eqref{eq:HamG}. As discussed at the end of Section \ref{ssec:geomObj}, the Hamilton geometry is reduced to the pseudo-Riemannian geometry of the spacetime. For the field equation this means $Q_{\mu\nu}=X_1{}_{\mu\nu}=X_2{}_{\mu\nu}=X_3{}_{\mu\nu} = 0$, since the Landsberg tensor vanishes as well as $\delta_\mu H$. The latter also implies that all other terms multiplied by $\alpha_2$ or $\alpha_3$ vanish. The only term that is non-vanishing in \eqref{eq:HamEOM} is
\begin{align}
    \tfrac{1}{2} \bar\partial^\mu \bar\partial^\nu \left(\frac{R_{\mu\nu}}{H}\right) - \frac{R}{H^2} = 0\,,
\end{align}
where, following the discussion at the end of Section \ref{ssec:geomObj}, $R_{\mu\nu} = R_{\mu\nu}[G]$ and $R=R[G]$ are the Ricci tensor and the Ricci scalar of the Levi-Civita connection of the space-time metric $G$. Applying the derivatives to $1/H$ we get
\begin{align}
   0 =  \tfrac{1}{2} R_{\mu\nu} \bar\partial^\nu \left(-\tfrac{2}{H^2}G^{\mu\sigma}p_\sigma\right) - \tfrac{R}{H^2} 
   = \tfrac{1}{2} R (-\tfrac{2}{H^2} G^{\mu\nu} + \tfrac{8}{H^3} G^{\mu\sigma}p_\sigma G^{\mu\rho}p_\rho)  - \tfrac{R}{H^2}
   = -  \tfrac{2}{H^2} R + \tfrac{4}{H^3} R^{\mu\nu}p_\mu p_\nu\,.
\end{align}
Multiplying this equation with $H^3/4 $ yields 
\begin{align}
    0 = -\frac{1}{2} H R + R^{\mu\nu}p_\mu p_\nu
    = \left( R^{\mu\nu} - \frac{1}{2} G^{\mu\nu}R\right)p_\mu p_\nu\,,
\end{align}
which is equivalent to the vanishing of the Einstein tensor $R^{\mu\nu} - \frac{1}{2} G^{\mu\nu}R = 0$. 

\subsection{The Einstein-Maxwell equations from one scalar field equation on phase space}\label{ssec:EinMax}
To derive the Einstein-Maxwell equations, we consider an Ansatz for the Hamiltonian of the form
\begin{align}\label{eq:HamEM}
    H_{EM}(x,p) = G^{\mu\nu}(x)\tilde p_\mu \tilde p_\nu := G^{\mu\nu}(x)(p_\mu - e A_\mu(x))(p_\nu - e A_\nu(x)) \,,
\end{align}
which describes the motion of a particle in curved spacetime in an electromagnetic field. The precise combination of particle momentum coordinate $p_\mu$ and electromagnetic potential $A_\mu$ in $\tilde p_\mu$ is necessary to ensure $U(1)$-invariance of the field equations, meaning that they only depend on $A_\mu$ through the electromagnetic field strength tensor $F_{\mu\nu} = \overset{\circ}{\nabla}_{[\mu} A_{\nu]}$. 

The Ansatz for the Hamiltonian brings some fortunate simplification when we derive the geometric objects, see Section \ref{ssec:geomObj}, of Hamilton geometry for this case:
\begin{itemize}
    \item The first $p$-derivative of $H$
    \begin{align}
        \bar\partial^\mu H = 2 G^{\mu\nu} \tilde p_\nu 
    \end{align}
    \item The Hamilton metric and the vertical connection coefficients
    \begin{align}
        g^{\mu\nu} = G^{\mu\nu}\,,\quad C_\sigma{}^{\mu\nu} = 0\,.
    \end{align}
    \item The non-linear (see \eqref{eq:NonLin}) connection coefficients
    \begin{align}\label{eq:nonlineGA}
        N_{\mu\nu} = \Gamma^\sigma{}_{\mu\nu} p_\sigma + e \overset{\circ}{\nabla}_{(\mu}A_{\nu)}\,,
    \end{align}
   where $\Gamma^\sigma{}_{\mu\nu}$ are the Christoffel symbols built from the spacetime metric $G$ and $\overset{\circ}{\nabla}$ is the Levi-Civita covariant derivative of $G$.
    \item The horizontal derivatives of $H$
    \begin{align}
        \delta_\mu H &= - e F_{\mu\nu} \tilde p^\nu \,,\label{eq:deltaH}\\
        \delta_\nu \delta_\mu H 
        &=  - e \partial_\nu F_{\mu\sigma} \tilde p^\sigma 
        - e F_{\mu\sigma}\left(-p^\beta \Gamma^\sigma{}_{\nu\beta}+ e \frac{1}{2}(\overset{\circ}{\nabla}_\nu A^\sigma + \overset{\circ}{\nabla}{}^\sigma A_\nu) - e \partial_\nu A^\sigma\right)\,,\\
      \nabla_\nu \delta_\mu H 
        &=  - e \overset{\circ}{\nabla}_\nu F_{\mu\sigma} \tilde p^\sigma 
        + \frac{e^2}{2} {F_{\mu}}^\sigma F_{\nu\sigma} \,.
    \end{align}
    
     \item The horizontal connection coefficients $H^\sigma{}_{\mu\nu}$ (see \eqref{eq:cartanCD}), the $p$-derivative of the non-linear connection coefficients and the Landsberg tensor (see \eqref{eq:Lands})
    \begin{align}
        H^\sigma{}_{\mu\nu} = \bar\partial^\sigma{}N_{\mu\nu} = \Gamma^\sigma{}_{\mu\nu} \Rightarrow S^\sigma{}_{\mu\nu} = 0\,.
    \end{align}
    
    \item The $4$-index curvature like tensor (see \eqref{eq:CartanCurv})
    \begin{align}
        \mathcal{R}^{\lambda}{}_{\rho\mu\nu} = R^{\lambda}{}_{\rho\mu\nu}
    \end{align}
    \item    
    The non-linear curvature tensor~\eqref{eq:NonLinCurv}
    \begin{equation}
        R_{\mu\nu\rho}\,=\,\tilde  p_\lambda {R^{\lambda}}_{\mu\nu\rho}+\frac{e}{2}\overset{\circ}{\nabla}{}_\mu F_{\rho\nu}\,,
    \label{eq:nl_curvature}
    \end{equation}
    plays a significant role in this process. Due to the symmetries of the non-linear curvature tensor~\cite{miron2001geometry} (note that the symmetries of the the nonlinear curvature tensor exploited below can be directly obtained from the definition~\eqref{eq:NonLinCurv} if the nonlinear coefficients are symmetric in their two indices)
    \begin{equation}\label{eq:nonlinBianchi}
        R_{[\mu\nu\rho]},=\,0\,,
    \end{equation}
    one obtains directly from Eq.~\eqref{eq:nl_curvature}
    \begin{equation}
        \,{R^\lambda}_{[\mu\nu\rho]}+\overset{\circ}{\nabla}{}_{\left[\rho\right.} F_{\left.\mu\nu\right]}\,=\overset{\circ}{\nabla}{}_{\left[\rho\right.} F_{\left.\mu\nu\right]}=\,0\,,
    \end{equation}
    where the first equality is due to the algebraic Bianchi identity of the Riemannian curvature tensor. Hence, the Faraday–Gauss law is easily recovered from the geometric structure of the Hamiltonian geometries.
\end{itemize}
Using these expressions, we can piece together the terms that make up the field equation \eqref{eq:HamEOM}:
\begin{itemize}
    \item The $Q$-tensor \eqref{eq:Qtensor}
    \begin{align}
        Q_{\mu\nu} 
        &= - \nabla_{\nu} \left(\tfrac{\delta_\mu H}{H^2} \right)
        =  2 \frac{\delta_\nu H \delta_\mu H}{H^3} - \frac{\nabla_\nu \delta_\mu H}{H^2}=\frac{2 e^2 F_{\mu \rho} F_{\nu \sigma} \tilde p^\rho \tilde p^\sigma}{H^3} 
        + \frac{e ( 2 \tilde p^\rho \overset{\circ}{\nabla}_{\nu}F_{\mu \rho} - e F_{\mu}{}^{\rho} F_{\nu \rho} )}{2 H^2}\,.
    \end{align}
       \item The $X_i^{\mu\nu}$ and $X^{\mu\nu}$ terms \eqref{eq:X1},\eqref{eq:X2},\eqref{eq:X3},
    \begin{align}
        X_1^{\mu\nu} &= 0\,,\\
        X_2^{\mu\nu} &= -\frac{2 e \tilde p_\sigma\left( \tilde p^\mu F^{\nu\sigma}+\tilde p^\nu F^{\mu\sigma}\right)}{H^2} \,,\\
        X_3^{\mu\nu} &= -2 X_2^{\mu\nu} \,,\\
        \Rightarrow X^{\mu\nu} &= \kappa X_2^{\mu\nu}\left(\alpha_2- 2\alpha_3\right)\,.
    \end{align}

    \item And last but not least the $X$-scalar \eqref{eq:defX}
    \begin{align}
        X &= \tfrac{1}{2}\bar\partial^\sigma \bar\partial^\lambda\left(    \bar\partial^\alpha (X_{(\lambda\sigma)}\delta_\alpha H  ) 
        - \tilde\nabla_\alpha(X_{(\lambda\sigma)}\bar\partial^\alpha H)
        + 2 X^{(\mu\nu)}g^H_{\sigma\mu} g^H_{\lambda\rho} \nabla_\nu \bar \partial^\rho H\right) 
        - \tilde\nabla_{\mu}\bar\partial^\rho(2X^{(\mu\nu)}g^H_{\nu\rho})\nonumber\\
        &= \left(\frac{8 e^2 (F_{\mu\nu} F^{\mu\nu} H - 4 F_{\mu}{}^{\nu} F_{\sigma\nu} \tilde p^\mu (\tilde p^\sigma))}{H^3} + \frac{16 e \tilde p^\mu \overset{\circ}{\nabla}_{\sigma}F_{\mu}{}^{\sigma}}{H^2} \right) \kappa (\alpha_2-2\alpha_3)\,.
    \end{align}
\end{itemize}

Combing all these terms to evaluate \eqref{eq:HamEOM} gives
\begin{align}
    0&=- \frac{2 \alpha_1 (H R - 2 R_{\mu\nu} \tilde{p}^{\mu} \tilde{p}^{\nu})}{H^3}\nonumber\\ 
    &+ \frac{\alpha_2 \kappa e \bigl(e F_{\mu\nu} F^{\mu\nu} H + 3 \tilde{p}^{\mu} (-2 e F_{\mu}{}^{\sigma} F_{\nu\sigma} \tilde{p}^{\nu} + H \overset{\circ}{\nabla}_{\nu }F_{\mu}{}^{\nu })\bigr)}{H^3} \nonumber\\
    &+ \frac{\alpha_3 \kappa e \bigl(-3 e F_{\mu\nu} F^{\mu\nu} H + 8 \tilde{p}^{\mu} (2 e F_{\mu}{}^{\sigma} F_{\nu \sigma} \tilde{p}^{\nu } -  H \overset{\circ}{\nabla}_{\nu }F_{\mu}{}^{\nu })\bigr)}{H^3}\,.
\end{align}
We immediately see that we recover vacuum general relativity, which we discussed in Section \ref{ssec:GR}, for $\alpha_2=\alpha_3=0$.

To further analyze the field equation, we multiply by $H^3$ and sort according to powers in $\tilde p$
\begin{align}
    0 &= \left[-2 \alpha_1 (G^{\mu\nu} R -2 R^{\mu\nu})
    + \kappa e^2 \left(2 (8 \alpha_3 - 3 \alpha_2)  F^{\mu\sigma} F^{\nu}{}_{\sigma} + (\alpha_2-3 \alpha_3 )  F_{\sigma \rho} F^{\sigma\rho} g^{\mu\nu}\right)\right]\tilde p_\mu \tilde p_\nu\nonumber\\
    &+ \left[\kappa e (3 \alpha_2 -8 \alpha_3)    g^{\mu\nu}  \overset{\circ}{\nabla}_{\rho}F^{\sigma\rho} \right]\tilde{p}_{\mu} \tilde{p}_{\nu} \tilde{p}_{\sigma}\,.
\end{align}
The thus rewritten field equations need to hold for all possible choices of $p_\mu$. Hence, each power of $\tilde p$ must vanish separately. The $\tilde p^3$ terms are clearly equivalent to the Maxwell vacuum equations
\begin{align}\label{eq:EinstMax1}
    g^{\mu\nu}  \overset{\circ}{\nabla}_{\rho}F^{\sigma\rho} = 0 \Leftrightarrow \overset{\circ}{\nabla}_{\rho}F^{\sigma\rho} = 0\,.
\end{align}
For the remaining $\tilde p^2$ Einstein like equations have to hold
\begin{align}\label{eq:EinstMax2}
    R^{\mu\nu} - \frac{1}{2}G^{\mu\nu} R   = - \frac{\kappa e^2}{4\alpha_1} \left(2 (8 \alpha_3 - 3 \alpha_2)  F^{\nu\sigma} F^{\mu}{}_{\sigma} + (\alpha_2-3 \alpha_3 )  F_{\sigma \rho} F^{\sigma\rho} g^{\mu\nu}\right)\,.
\end{align}
To identify the right-hand side with the energy-momentum tensor that one usually finds from the variation of the Maxwell action on curved spacetime with respect to the metric we need to choose the theory parameters according to the following constraints
\begin{align}
     2(8\alpha_3 - 3 \alpha_2) &= \Xi \,, \\
    (\alpha_2-3 \alpha_3 ) &= -\frac{1}{4}\Xi\,,
\end{align}
where $\Xi$ is a constant. Thus, the covariant divergence of the right-hand side of Eq.~\eqref{eq:EinstMax2} vanishes, as does the trace. These constraint equations imply that
\begin{align}
    \alpha_2= 2 \alpha_3\,, 
\end{align}
which makes the equations \eqref{eq:EinstMax2}
\begin{align}\label{eq:EinstMax3}
    R^{\mu\nu} - \frac{1}{2}G^{\mu\nu} R   = - \frac{\kappa e^2 \alpha_3}{\alpha_1} \left(  F^{\nu\sigma} F^{\mu}{}_{\sigma} -  \frac{1}{4}  F_{\sigma \rho} F^{\sigma\rho} g^{\mu\nu}\right)\,.
\end{align}
Hence, by choosing $e^2 \alpha_3/\alpha_1 = -1$, we obtain the Einstein–Maxwell equations from the general Hamiltonian field equation on the cotangent bundle \eqref{eq:HamEOM}.

This derivation demonstrates that the geometry of Hamiltonian space $(M,H_{EM})$ encodes gravity and electromagnetism into the Hamilton geometry of spacetime, with the help of the Hamiltonian scalar field equation.

\section{Discussion}\label{sec:discussion}
In this work, we proposed a novel approach to unify gravitational and electromagnetic fields using cotangent bundle geometries based on a cotangent bundle scalar field, the point-particle Hamiltonian. In contrast to other approaches, the electromagnetic 4-potential and spacetime metric are encoded in the Hamilton function, and their interaction emerges in the non-linear connection coefficients \eqref{eq:nonlineGA}, such that the electromagnetic force and gravity appear directly in the geometrical ingredients of the cotangent bundle, that is, in the geometry of the one-particle phase space of spacetime. 

We constructed an action principle, \eqref{eq:act}, for Hamilton functions as scalar fields in phase space and derived field equations, \eqref{eq:HamEOM}, that determine the Hamiltonian. Remarkably, we can fix the coupling parameters in the action so that these field equations reduce to the coupled Einstein–Maxwell equations when using the Hamiltonian describing the motion of a charged particle in a curved spacetime embedded in an electromagnetic field, as demonstrated in Section \ref{ssec:EinMax}. In addition, the  Faraday--Gauss law of electromagnetism emerges from purely geometrical arguments through the vanishing of the total antisymmetrization of the nonlinear curvature tensor \eqref{eq:nonlinBianchi}. In the limit of a vanishing charge parameter for the point particle Hamiltonian, we showed that one consistently recovers the vacuum Einstein equations in Section \ref{ssec:GR}.

Thus, the framework we constructed encodes gravity and electromagnetism simultaneously into the geometry of phase space and can  be viewed as an extension of the understanding of physical interactions through geometry that was started by general relativity, where gravity is understood as the effect of the geometry of spacetime. One striking feature of our formulation of a unified field theory in phase space is that it does not require additional fields beyond a scalar field or the point particle Hamiltonian, nor does it require additional spacetime dimensions.

Having successfully unified electromagnetic and gravitational interactions into the geometry of phase space, the immediate next question is: what can we learn about the additional field from the cotangent bundle field equations? Considering a a general polynomial Hamiltonian of the form
\begin{align}
    H = \sum_{i=0}^N G^{a_1...a_i}(x)p_{a_1}...p_{a_i} 
    = G(x) + G^{a}(x)p_a + G^{ab}(x) p_a p_b + G^{abc}(x) p_a p_b p_c + ...\,,
\end{align}
the field equations \eqref{eq:HamEOM} will contain terms of different powers of the particle momenta $p$, and each power will lead to a field equation on spacetime, which together determine the spacetime fields $G^{a_1...a_i}(x), i=0,...N$. This research direction directly leads to a new perspective on the derivation of field equations for higher-spin fields, which are usually approached from different angles \cite{Singh:1974qz,Fronsdal:1978rb,Sagnotti:2011jdy,Sorokin:2004ie,Ponomarev:2022vjb,Bekaert:2004qos,Giombi:2016ejx,Tomasiello:2024jyu}. 

Another question is whether the scalar field action we presented here is (up to boundary terms) the unique viable action for a scalar field on phase space such that it reproduces the Einstein-Maxwell equations, which is most likely not the case. A systematic classification of scalar field actions in phase space, which can recover known physics and field equations for higher-spin fields, is a work in progress. Therefore, an essential ingredient to be able to perform such a study is the geometric framework of Hamiltonian geometry, since it allows us to systematically identify the phase space scalars that are invariant under diffeomorphisms of spacetime.

Moreover, the results of this paper lay the groundwork for generalizing the Einstein equations to provide dynamics to point particle Hamiltonians that are relevant in the context of quantum gravity phenomenology~\cite{Addazi:2021xuf}, and emerge in scenarios involving Lorentz invariance violation (LIV)~\cite{Colladay:1998fq,Kostelecky:2011qz} or deformed special relativity (DSR)~\cite{Amelino-Camelia2002b,AmelinoCamelia:2001vy,AmelinoCamelia:2008qg}. Our findings indicate that if one changes the Hamiltonian, one changes the electromagnetic and gravitational dynamics simultaneously. If this is the case or not, can be derived from the phase space field equation. The resulting dynamics can then be compared to earlier papers on gravity and electromagnetism in LIV~\cite{Lammerzahl:2005jw,Xiao:2010yx,Casana:2011du,Chang:2012ks} (including approaches with velocity-dependent geometries~\cite{Crisan:2020krc}) DSR literature~\cite{Takka:2019him}. 

Regarding the latter, it has recently been shown~\cite{Relancio:2020rys,Pfeifer:2021tas} that a more complex structure than that of Hamilton spaces (so-called generalized Hamilton spaces) is needed to describe the deformed relativistic kinematics of DSR. These geometries are derived not only from a scalar field in phase space, but also from a metric in momentum space, and the point-particle Hamiltonian is derived from this momentum space metric as  a squared geometric distance~\cite{Relancio:2024nud}. To extend the Einstein equations to this scenario and determine the momentum space metric dynamically, further extensions of this work are required.

In conclusion, this work provides a new perspective on the geometric foundations of field theories and offers a promising framework for unifying fundamental interactions from the perspective of cotangent bundle geometry. Beyond that, it offers a new way to generalize dynamical equations for geometries that are applicable to quantum gravity phenomenology.

\section*{Acknowledgements}
CP acknowledges the financial support by the excellence cluster QuantumFrontiers of the German Research Foundation (Deutsche Forschungsgemeinschaft, DFG) under Germany's Excellence Strategy -- EXC-2123 QuantumFrontiers -- 390837967 and was funded by the Deutsche Forschungsgemeinschaft (DFG, German Research Foundation) - Project Number 420243324. JJR acknowledge partial support from the grant PID2023-148373NB-I00 funded by MCIN/ AEI / 10.13039/501100011033/FEDER -- UE, and the Q-CAYLE Project funded by the Regional Government of Castilla y León (Junta de Castilla y León) and the Ministry of Science and Innovation MICIN through NextGenerationEU (PRTR C17.I1). The authors would like to acknowledge networking support by the COST Action CA23130.


\begin{thebibliography}{49}%
\makeatletter
\providecommand \@ifxundefined [1]{%
 \@ifx{#1\undefined}
}%
\providecommand \@ifnum [1]{%
 \ifnum #1\expandafter \@firstoftwo
 \else \expandafter \@secondoftwo
 \fi
}%
\providecommand \@ifx [1]{%
 \ifx #1\expandafter \@firstoftwo
 \else \expandafter \@secondoftwo
 \fi
}%
\providecommand \natexlab [1]{#1}%
\providecommand \enquote  [1]{``#1''}%
\providecommand \bibnamefont  [1]{#1}%
\providecommand \bibfnamefont [1]{#1}%
\providecommand \citenamefont [1]{#1}%
\providecommand \href@noop [0]{\@secondoftwo}%
\providecommand \href [0]{\begingroup \@sanitize@url \@href}%
\providecommand \@href[1]{\@@startlink{#1}\@@href}%
\providecommand \@@href[1]{\endgroup#1\@@endlink}%
\providecommand \@sanitize@url [0]{\catcode `\\12\catcode `\$12\catcode
  `\&12\catcode `\#12\catcode `\^12\catcode `\_12\catcode `\%12\relax}%
\providecommand \@@startlink[1]{}%
\providecommand \@@endlink[0]{}%
\providecommand \url  [0]{\begingroup\@sanitize@url \@url }%
\providecommand \@url [1]{\endgroup\@href {#1}{\urlprefix }}%
\providecommand \urlprefix  [0]{URL }%
\providecommand \Eprint [0]{\href }%
\providecommand \doibase [0]{https://doi.org/}%
\providecommand \selectlanguage [0]{\@gobble}%
\providecommand \bibinfo  [0]{\@secondoftwo}%
\providecommand \bibfield  [0]{\@secondoftwo}%
\providecommand \translation [1]{[#1]}%
\providecommand \BibitemOpen [0]{}%
\providecommand \bibitemStop [0]{}%
\providecommand \bibitemNoStop [0]{.\EOS\space}%
\providecommand \EOS [0]{\spacefactor3000\relax}%
\providecommand \BibitemShut  [1]{\csname bibitem#1\endcsname}%
\let\auto@bib@innerbib\@empty
\bibitem [{\citenamefont {Eddington}(1921)}]{Eddington1921}%
  \BibitemOpen
  \bibfield  {author} {\bibinfo {author} {\bibfnamefont {A.~S.}\ \bibnamefont
  {Eddington}},\ }\bibfield  {title} {\bibinfo {title} {{A generalisation of
  Weyl's theory of the electromagnetic and gravitational fields}},\ }\href
  {https://doi.org/10.1098/rspa.1921.0027} {\bibfield  {journal} {\bibinfo
  {journal} {Proceedings of the Royal Society of London. Series A, Containing
  Papers of a Mathematical and Physical Character}\ }\textbf {\bibinfo {volume}
  {99}},\ \bibinfo {pages} {104–122} (\bibinfo {year} {1921})}\BibitemShut
  {NoStop}%
\bibitem [{\citenamefont {Weyl}(1922)}]{Weyl1922}%
  \BibitemOpen
  \bibfield  {author} {\bibinfo {author} {\bibfnamefont {H.}~\bibnamefont
  {Weyl}},\ }\bibinfo {title} {{Gravitation und Elektrizit{\"a}t}},\ in\ \href
  {https://doi.org/10.1007/978-3-663-16170-7_11} {\emph {\bibinfo {booktitle}
  {Das Relativit{\"a}tsprinzip: Eine Sammlung von Abhandlungen}}}\ (\bibinfo
  {publisher} {Vieweg+Teubner Verlag},\ \bibinfo {address} {Wiesbaden},\
  \bibinfo {year} {1922})\ pp.\ \bibinfo {pages} {147--159}\BibitemShut
  {NoStop}%
\bibitem [{\citenamefont {Cartan}(1924)}]{Cartan:1924yea}%
  \BibitemOpen
  \bibfield  {author} {\bibinfo {author} {\bibfnamefont {E.}~\bibnamefont
  {Cartan}},\ }\bibfield  {title} {\bibinfo {title} {{Sur les vari{\'e}t{\'e}s
  {\`a} connexion affine et la th{\'e}orie de la relativit{\'e}
  g{\'e}n{\'e}ralis{\'e}e. (premi{\`e}re partie) (Suite).}},\ }\href@noop {}
  {\bibfield  {journal} {\bibinfo  {journal} {Annales Sci. Ecole Norm. Sup.}\
  }\textbf {\bibinfo {volume} {41}},\ \bibinfo {pages} {1} (\bibinfo {year}
  {1924})}\BibitemShut {NoStop}%
\bibitem [{\citenamefont {Einstein}(1929)}]{Einstein1929}%
  \BibitemOpen
  \bibfield  {author} {\bibinfo {author} {\bibfnamefont {A.}~\bibnamefont
  {Einstein}},\ }\bibfield  {title} {\bibinfo {title} {Zur einheitlichen
  feldtheorie},\ }\href@noop {} {\bibfield  {journal} {\bibinfo  {journal}
  {Sitzungsberichte der Preussischen Akademie der Wissenschaften zu Berlin}\ ,\
  \bibinfo {pages} {2}} (\bibinfo {year} {1929})}\BibitemShut {NoStop}%
\bibitem [{\citenamefont {Ringermacher}(1994)}]{Ringermacher_1994}%
  \BibitemOpen
  \bibfield  {author} {\bibinfo {author} {\bibfnamefont {H.~I.}\ \bibnamefont
  {Ringermacher}},\ }\bibfield  {title} {\bibinfo {title} {An electrodynamic
  connection},\ }\href {https://doi.org/10.1088/0264-9381/11/9/018} {\bibfield
  {journal} {\bibinfo  {journal} {Classical and Quantum Gravity}\ }\textbf
  {\bibinfo {volume} {11}},\ \bibinfo {pages} {2383} (\bibinfo {year}
  {1994})}\BibitemShut {NoStop}%
\bibitem [{\citenamefont {Kohler}(2000)}]{Kohler:1998ah}%
  \BibitemOpen
  \bibfield  {author} {\bibinfo {author} {\bibfnamefont {C.}~\bibnamefont
  {Kohler}},\ }\bibfield  {title} {\bibinfo {title} {{Einstein-Cartan-Maxwell
  theory with scalar field through a five-dimensional unification}},\ }\href
  {https://doi.org/10.1142/S0217751X00000562} {\bibfield  {journal} {\bibinfo
  {journal} {Int. J. Mod. Phys. A}\ }\textbf {\bibinfo {volume} {15}},\
  \bibinfo {pages} {1235} (\bibinfo {year} {2000})},\ \Eprint
  {https://arxiv.org/abs/gr-qc/9808004} {arXiv:gr-qc/9808004} \BibitemShut
  {NoStop}%
\bibitem [{\citenamefont {Janssen}\ and\ \citenamefont
  {Jim{\'e}nez-Cano}(2018)}]{Janssen:2018exh}%
  \BibitemOpen
  \bibfield  {author} {\bibinfo {author} {\bibfnamefont {B.}~\bibnamefont
  {Janssen}}\ and\ \bibinfo {author} {\bibfnamefont {A.}~\bibnamefont
  {Jim{\'e}nez-Cano}},\ }\bibfield  {title} {\bibinfo {title} {{Projective
  symmetries and induced electromagnetism in metric-affine gravity}},\ }\href
  {https://doi.org/10.1016/j.physletb.2018.10.032} {\bibfield  {journal}
  {\bibinfo  {journal} {Phys. Lett. B}\ }\textbf {\bibinfo {volume} {786}},\
  \bibinfo {pages} {462} (\bibinfo {year} {2018})},\ \Eprint
  {https://arxiv.org/abs/1807.10168} {arXiv:1807.10168 [gr-qc]} \BibitemShut
  {NoStop}%
\bibitem [{\citenamefont {Beach}(2024)}]{BEACH2024169661}%
  \BibitemOpen
  \bibfield  {author} {\bibinfo {author} {\bibfnamefont {R.~J.}\ \bibnamefont
  {Beach}},\ }\bibfield  {title} {\bibinfo {title} {The geometrization of
  maxwell's equations and the emergence of gravity and antimatter},\ }\href
  {https://doi.org/https://doi.org/10.1016/j.aop.2024.169661} {\bibfield
  {journal} {\bibinfo  {journal} {Annals of Physics}\ }\textbf {\bibinfo
  {volume} {465}},\ \bibinfo {pages} {169661} (\bibinfo {year}
  {2024})}\BibitemShut {NoStop}%
\bibitem [{\citenamefont {Hehl}\ \emph {et~al.}(1995)\citenamefont {Hehl},
  \citenamefont {McCrea}, \citenamefont {Mielke},\ and\ \citenamefont
  {Ne'eman}}]{Hehl:1994ue}%
  \BibitemOpen
  \bibfield  {author} {\bibinfo {author} {\bibfnamefont {F.~W.}\ \bibnamefont
  {Hehl}}, \bibinfo {author} {\bibfnamefont {J.~D.}\ \bibnamefont {McCrea}},
  \bibinfo {author} {\bibfnamefont {E.~W.}\ \bibnamefont {Mielke}},\ and\
  \bibinfo {author} {\bibfnamefont {Y.}~\bibnamefont {Ne'eman}},\ }\bibfield
  {title} {\bibinfo {title} {{Metric affine gauge theory of gravity: Field
  equations, Noether identities, world spinors, and breaking of dilation
  invariance}},\ }\href {https://doi.org/10.1016/0370-1573(94)00111-F}
  {\bibfield  {journal} {\bibinfo  {journal} {Phys. Rept.}\ }\textbf {\bibinfo
  {volume} {258}},\ \bibinfo {pages} {1} (\bibinfo {year} {1995})},\ \Eprint
  {https://arxiv.org/abs/gr-qc/9402012} {arXiv:gr-qc/9402012} \BibitemShut
  {NoStop}%
\bibitem [{\citenamefont {Vizgin}(1994)}]{Vizgin1994}%
  \BibitemOpen
  \bibfield  {author} {\bibinfo {author} {\bibfnamefont {V.~P.}\ \bibnamefont
  {Vizgin}},\ }\href {https://doi.org/10.1007/978-3-0348-8528-8} {\emph
  {\bibinfo {title} {Unified Field Theories in the First Third of the 20th
  Century}}}\ (\bibinfo  {publisher} {Birkhäuser},\ \bibinfo {year}
  {1994})\BibitemShut {NoStop}%
\bibitem [{\citenamefont {Goenner}(2004)}]{Goenner:2004se}%
  \BibitemOpen
  \bibfield  {author} {\bibinfo {author} {\bibfnamefont {H.~F.~M.}\
  \bibnamefont {Goenner}},\ }\bibfield  {title} {\bibinfo {title} {{On the
  history of unified field theories}},\ }\href
  {https://doi.org/10.12942/lrr-2004-2} {\bibfield  {journal} {\bibinfo
  {journal} {Living Rev. Rel.}\ }\textbf {\bibinfo {volume} {7}},\ \bibinfo
  {pages} {2} (\bibinfo {year} {2004})}\BibitemShut {NoStop}%
\bibitem [{\citenamefont {{Kaluza}}(1921)}]{Kaluza_1921}%
  \BibitemOpen
  \bibfield  {author} {\bibinfo {author} {\bibfnamefont {T.}~\bibnamefont
  {{Kaluza}}},\ }\bibfield  {title} {\bibinfo {title} {{Zum Unit{\"a}tsproblem
  der Physik}},\ }\href@noop {} {\bibfield  {journal} {\bibinfo  {journal}
  {Sitzungsberichte der K{\"o}niglich Preussischen Akademie der
  Wissenschaften}\ ,\ \bibinfo {pages} {966}} (\bibinfo {year}
  {1921})}\BibitemShut {NoStop}%
\bibitem [{\citenamefont {Klein}(1926)}]{Klein:1926tv}%
  \BibitemOpen
  \bibfield  {author} {\bibinfo {author} {\bibfnamefont {O.}~\bibnamefont
  {Klein}},\ }\bibfield  {title} {\bibinfo {title} {{Quantum Theory and
  Five-Dimensional Theory of Relativity. (In German and English)}},\ }\href
  {https://doi.org/10.1007/BF01397481} {\bibfield  {journal} {\bibinfo
  {journal} {Z. Phys.}\ }\textbf {\bibinfo {volume} {37}},\ \bibinfo {pages}
  {895} (\bibinfo {year} {1926})}\BibitemShut {NoStop}%
\bibitem [{\citenamefont {Polchinski}(1998)}]{Polchinski_1998}%
  \BibitemOpen
  \bibfield  {author} {\bibinfo {author} {\bibfnamefont {J.}~\bibnamefont
  {Polchinski}},\ }\href@noop {} {\emph {\bibinfo {title} {String Theory}}},\
  Cambridge Monographs on Mathematical Physics\ (\bibinfo  {publisher}
  {Cambridge University Press},\ \bibinfo {year} {1998})\BibitemShut {NoStop}%
\bibitem [{\citenamefont {Mukhi}(2011)}]{Mukhi:2011zz}%
  \BibitemOpen
  \bibfield  {author} {\bibinfo {author} {\bibfnamefont {S.}~\bibnamefont
  {Mukhi}},\ }\bibfield  {title} {\bibinfo {title} {{String theory: a
  perspective over the last 25 years}},\ }\href
  {https://doi.org/10.1088/0264-9381/28/15/153001} {\bibfield  {journal}
  {\bibinfo  {journal} {Class. Quant. Grav.}\ }\textbf {\bibinfo {volume}
  {28}},\ \bibinfo {pages} {153001} (\bibinfo {year} {2011})},\ \Eprint
  {https://arxiv.org/abs/1110.2569} {arXiv:1110.2569 [physics.pop-ph]}
  \BibitemShut {NoStop}%
\bibitem [{\citenamefont {Marchesano}\ \emph {et~al.}(2024)\citenamefont
  {Marchesano}, \citenamefont {Shiu},\ and\ \citenamefont
  {Weigand}}]{Marchesano:2024gul}%
  \BibitemOpen
  \bibfield  {author} {\bibinfo {author} {\bibfnamefont {F.}~\bibnamefont
  {Marchesano}}, \bibinfo {author} {\bibfnamefont {G.}~\bibnamefont {Shiu}},\
  and\ \bibinfo {author} {\bibfnamefont {T.}~\bibnamefont {Weigand}},\
  }\bibfield  {title} {\bibinfo {title} {{The Standard Model from String
  Theory: What Have We Learned?}},\ }\href
  {https://doi.org/10.1146/annurev-nucl-102622-012235} {\bibfield  {journal}
  {\bibinfo  {journal} {Ann. Rev. Nucl. Part. Sci.}\ }\textbf {\bibinfo
  {volume} {74}},\ \bibinfo {pages} {113} (\bibinfo {year} {2024})},\ \Eprint
  {https://arxiv.org/abs/2401.01939} {arXiv:2401.01939 [hep-th]} \BibitemShut
  {NoStop}%
\bibitem [{\citenamefont {Miron}\ \emph {et~al.}(2001)\citenamefont {Miron},
  \citenamefont {Hrimiuc}, \citenamefont {Shimada},\ and\ \citenamefont
  {Sabau}}]{miron2001geometry}%
  \BibitemOpen
  \bibfield  {author} {\bibinfo {author} {\bibfnamefont {R.}~\bibnamefont
  {Miron}}, \bibinfo {author} {\bibfnamefont {D.}~\bibnamefont {Hrimiuc}},
  \bibinfo {author} {\bibfnamefont {H.}~\bibnamefont {Shimada}},\ and\ \bibinfo
  {author} {\bibfnamefont {S.}~\bibnamefont {Sabau}},\ }\href
  {https://books.google.es/books?id=l3JNMzL14SAC} {\emph {\bibinfo {title} {The
  Geometry of Hamilton and Lagrange Spaces}}},\ Fundamental Theories of
  Physics\ (\bibinfo  {publisher} {Springer},\ \bibinfo {year}
  {2001})\BibitemShut {NoStop}%
\bibitem [{\citenamefont {{Miron}}(2012)}]{Miron_HamiltonGeometry_2012}%
  \BibitemOpen
  \bibfield  {author} {\bibinfo {author} {\bibfnamefont {R.}~\bibnamefont
  {{Miron}}},\ }\bibfield  {title} {\bibinfo {title} {{Lagrangian and
  Hamiltonian Geometries. Applications to Analytical Mechanics}},\ }\href@noop
  {} {\bibfield  {journal} {\bibinfo  {journal} {arXiv e-prints}\ ,\ \bibinfo
  {eid} {arXiv:1203.4101}} (\bibinfo {year} {2012})},\ \Eprint
  {https://arxiv.org/abs/1203.4101} {arXiv:1203.4101 [math.DG]} \BibitemShut
  {NoStop}%
\bibitem [{\citenamefont {Rund}(2012)}]{Rund2012}%
  \BibitemOpen
  \bibfield  {author} {\bibinfo {author} {\bibfnamefont {H.}~\bibnamefont
  {Rund}},\ }\href@noop {} {\emph {\bibinfo {title} {The differential geometry
  of Finsler spaces}}},\ Vol.\ \bibinfo {volume} {101}\ (\bibinfo  {publisher}
  {Springer Science \& Business Media},\ \bibinfo {year} {2012})\BibitemShut
  {NoStop}%
\bibitem [{\citenamefont {Miron}\ and\ \citenamefont
  {Bucataru}(2007)}]{Bucataru}%
  \BibitemOpen
  \bibfield  {author} {\bibinfo {author} {\bibfnamefont {R.}~\bibnamefont
  {Miron}}\ and\ \bibinfo {author} {\bibfnamefont {I.}~\bibnamefont
  {Bucataru}},\ }\href@noop {} {\emph {\bibinfo {title} {Finsler Lagrange
  geometry}}}\ (\bibinfo  {publisher} {Editura Academiei Romane},\ \bibinfo
  {year} {2007})\BibitemShut {NoStop}%
\bibitem [{\citenamefont {Bao}\ \emph {et~al.}(2000)\citenamefont {Bao},
  \citenamefont {Chern},\ and\ \citenamefont {Shen}}]{BCS}%
  \BibitemOpen
  \bibfield  {author} {\bibinfo {author} {\bibfnamefont {D.}~\bibnamefont
  {Bao}}, \bibinfo {author} {\bibfnamefont {S.-S.}\ \bibnamefont {Chern}},\
  and\ \bibinfo {author} {\bibfnamefont {Z.}~\bibnamefont {Shen}},\ }\href@noop
  {} {\emph {\bibinfo {title} {{An Introduction to Finsler-Riemann
  Geometry}}}}\ (\bibinfo  {publisher} {Springer, New York},\ \bibinfo {year}
  {2000})\BibitemShut {NoStop}%
\bibitem [{\citenamefont {Hohmann}\ \emph {et~al.}(2022)\citenamefont
  {Hohmann}, \citenamefont {Pfeifer},\ and\ \citenamefont
  {Voicu}}]{Hohmann:2021zbt}%
  \BibitemOpen
  \bibfield  {author} {\bibinfo {author} {\bibfnamefont {M.}~\bibnamefont
  {Hohmann}}, \bibinfo {author} {\bibfnamefont {C.}~\bibnamefont {Pfeifer}},\
  and\ \bibinfo {author} {\bibfnamefont {N.}~\bibnamefont {Voicu}},\ }\bibfield
   {title} {\bibinfo {title} {{Mathematical foundations for field theories on
  Finsler spacetimes}},\ }\href {https://doi.org/10.1063/5.0065944} {\bibfield
  {journal} {\bibinfo  {journal} {J. Math. Phys.}\ }\textbf {\bibinfo {volume}
  {63}},\ \bibinfo {pages} {032503} (\bibinfo {year} {2022})},\ \Eprint
  {https://arxiv.org/abs/2106.14965} {arXiv:2106.14965 [math-ph]} \BibitemShut
  {NoStop}%
\bibitem [{\citenamefont {Barcaroli}\ \emph {et~al.}(2015)\citenamefont
  {Barcaroli}, \citenamefont {Brunkhorst}, \citenamefont {Gubitosi},
  \citenamefont {Loret},\ and\ \citenamefont {Pfeifer}}]{Barcaroli:2015xda}%
  \BibitemOpen
  \bibfield  {author} {\bibinfo {author} {\bibfnamefont {L.}~\bibnamefont
  {Barcaroli}}, \bibinfo {author} {\bibfnamefont {L.~K.}\ \bibnamefont
  {Brunkhorst}}, \bibinfo {author} {\bibfnamefont {G.}~\bibnamefont
  {Gubitosi}}, \bibinfo {author} {\bibfnamefont {N.}~\bibnamefont {Loret}},\
  and\ \bibinfo {author} {\bibfnamefont {C.}~\bibnamefont {Pfeifer}},\
  }\bibfield  {title} {\bibinfo {title} {{Hamilton geometry: Phase space
  geometry from modified dispersion relations}},\ }\href
  {https://doi.org/10.1103/PhysRevD.92.084053} {\bibfield  {journal} {\bibinfo
  {journal} {Phys. Rev. D}\ }\textbf {\bibinfo {volume} {92}},\ \bibinfo
  {pages} {084053} (\bibinfo {year} {2015})},\ \Eprint
  {https://arxiv.org/abs/1507.00922} {arXiv:1507.00922 [gr-qc]} \BibitemShut
  {NoStop}%
\bibitem [{\citenamefont {Pfeifer}\ and\ \citenamefont
  {Relancio}(2022)}]{Pfeifer:2021tas}%
  \BibitemOpen
  \bibfield  {author} {\bibinfo {author} {\bibfnamefont {C.}~\bibnamefont
  {Pfeifer}}\ and\ \bibinfo {author} {\bibfnamefont {J.~J.}\ \bibnamefont
  {Relancio}},\ }\bibfield  {title} {\bibinfo {title} {{Deformed relativistic
  kinematics on curved spacetime: a geometric approach}},\ }\href
  {https://doi.org/10.1140/epjc/s10052-022-10066-w} {\bibfield  {journal}
  {\bibinfo  {journal} {Eur. Phys. J. C}\ }\textbf {\bibinfo {volume} {82}},\
  \bibinfo {pages} {150} (\bibinfo {year} {2022})},\ \Eprint
  {https://arxiv.org/abs/2103.16626} {arXiv:2103.16626 [gr-qc]} \BibitemShut
  {NoStop}%
\bibitem [{\citenamefont {Relancio}(2022)}]{Relancio:2022mia}%
  \BibitemOpen
  \bibfield  {author} {\bibinfo {author} {\bibfnamefont {J.~J.}\ \bibnamefont
  {Relancio}},\ }\bibfield  {title} {\bibinfo {title} {{Relativistic deformed
  kinematics: From flat to curved spacetimes}},\ }\href
  {https://doi.org/10.1142/S0219887822300045} {\bibfield  {journal} {\bibinfo
  {journal} {Int. J. Geom. Meth. Mod. Phys.}\ }\textbf {\bibinfo {volume}
  {19}},\ \bibinfo {pages} {2230004} (\bibinfo {year} {2022})},\ \Eprint
  {https://arxiv.org/abs/2207.08471} {arXiv:2207.08471 [gr-qc]} \BibitemShut
  {NoStop}%
\bibitem [{\citenamefont {Perlick}(2000)}]{Perlick2000RayOptics}%
  \BibitemOpen
  \bibfield  {author} {\bibinfo {author} {\bibfnamefont {V.}~\bibnamefont
  {Perlick}},\ }\href {https://doi.org/10.1007/3-540-46662-2} {\emph {\bibinfo
  {title} {Ray Optics, Fermat’s Principle, and Applications to General
  Relativity}}},\ \bibinfo {edition} {1st}\ ed.,\ \bibinfo {series} {Lecture
  Notes in Physics Monographs}, Vol.~\bibinfo {volume} {61}\ (\bibinfo
  {publisher} {Springer Berlin Heidelberg},\ \bibinfo {address} {Berlin,
  Heidelberg},\ \bibinfo {year} {2000})\ pp.\ \bibinfo {pages} {x,
  220}\BibitemShut {NoStop}%
\bibitem [{\citenamefont {Relancio}\ and\ \citenamefont
  {Liberati}(2021)}]{Relancio:2020rys}%
  \BibitemOpen
  \bibfield  {author} {\bibinfo {author} {\bibfnamefont {J.~J.}\ \bibnamefont
  {Relancio}}\ and\ \bibinfo {author} {\bibfnamefont {S.}~\bibnamefont
  {Liberati}},\ }\bibfield  {title} {\bibinfo {title} {{Towards a geometrical
  interpretation of rainbow geometries}},\ }\href
  {https://doi.org/10.1088/1361-6382/ac05d7} {\bibfield  {journal} {\bibinfo
  {journal} {Class. Quant. Grav.}\ }\textbf {\bibinfo {volume} {38}},\ \bibinfo
  {pages} {135028} (\bibinfo {year} {2021})},\ \Eprint
  {https://arxiv.org/abs/2010.15734} {arXiv:2010.15734 [gr-qc]} \BibitemShut
  {NoStop}%
\bibitem [{\citenamefont {Hohmann}\ \emph {et~al.}(2019)\citenamefont
  {Hohmann}, \citenamefont {Pfeifer},\ and\ \citenamefont
  {Voicu}}]{Hohmann:2018rpp}%
  \BibitemOpen
  \bibfield  {author} {\bibinfo {author} {\bibfnamefont {M.}~\bibnamefont
  {Hohmann}}, \bibinfo {author} {\bibfnamefont {C.}~\bibnamefont {Pfeifer}},\
  and\ \bibinfo {author} {\bibfnamefont {N.}~\bibnamefont {Voicu}},\ }\bibfield
   {title} {\bibinfo {title} {{Finsler gravity action from variational
  completion}},\ }\href {https://doi.org/10.1103/PhysRevD.100.064035}
  {\bibfield  {journal} {\bibinfo  {journal} {Phys. Rev. D}\ }\textbf {\bibinfo
  {volume} {100}},\ \bibinfo {pages} {064035} (\bibinfo {year} {2019})},\
  \Eprint {https://arxiv.org/abs/1812.11161} {arXiv:1812.11161 [gr-qc]}
  \BibitemShut {NoStop}%
\bibitem [{\citenamefont {Singh}\ and\ \citenamefont
  {Hagen}(1974)}]{Singh:1974qz}%
  \BibitemOpen
  \bibfield  {author} {\bibinfo {author} {\bibfnamefont {L.~P.~S.}\
  \bibnamefont {Singh}}\ and\ \bibinfo {author} {\bibfnamefont {C.~R.}\
  \bibnamefont {Hagen}},\ }\bibfield  {title} {\bibinfo {title} {{Lagrangian
  formulation for arbitrary spin. 1. The boson case}},\ }\href
  {https://doi.org/10.1103/PhysRevD.9.898} {\bibfield  {journal} {\bibinfo
  {journal} {Phys. Rev. D}\ }\textbf {\bibinfo {volume} {9}},\ \bibinfo {pages}
  {898} (\bibinfo {year} {1974})}\BibitemShut {NoStop}%
\bibitem [{\citenamefont {Fronsdal}(1978)}]{Fronsdal:1978rb}%
  \BibitemOpen
  \bibfield  {author} {\bibinfo {author} {\bibfnamefont {C.}~\bibnamefont
  {Fronsdal}},\ }\bibfield  {title} {\bibinfo {title} {{Massless Fields with
  Integer Spin}},\ }\href {https://doi.org/10.1103/PhysRevD.18.3624} {\bibfield
   {journal} {\bibinfo  {journal} {Phys. Rev. D}\ }\textbf {\bibinfo {volume}
  {18}},\ \bibinfo {pages} {3624} (\bibinfo {year} {1978})}\BibitemShut
  {NoStop}%
\bibitem [{\citenamefont {Sagnotti}(2013)}]{Sagnotti:2011jdy}%
  \BibitemOpen
  \bibfield  {author} {\bibinfo {author} {\bibfnamefont {A.}~\bibnamefont
  {Sagnotti}},\ }\bibfield  {title} {\bibinfo {title} {{Notes on Strings and
  Higher Spins}},\ }\href {https://doi.org/10.1088/1751-8113/46/21/214006}
  {\bibfield  {journal} {\bibinfo  {journal} {J. Phys. A}\ }\textbf {\bibinfo
  {volume} {46}},\ \bibinfo {pages} {214006} (\bibinfo {year} {2013})},\
  \Eprint {https://arxiv.org/abs/1112.4285} {arXiv:1112.4285 [hep-th]}
  \BibitemShut {NoStop}%
\bibitem [{\citenamefont {Sorokin}(2005)}]{Sorokin:2004ie}%
  \BibitemOpen
  \bibfield  {author} {\bibinfo {author} {\bibfnamefont {D.}~\bibnamefont
  {Sorokin}},\ }\bibfield  {title} {\bibinfo {title} {{Introduction to the
  classical theory of higher spins}},\ }\href
  {https://doi.org/10.1063/1.1923335} {\bibfield  {journal} {\bibinfo
  {journal} {AIP Conf. Proc.}\ }\textbf {\bibinfo {volume} {767}},\ \bibinfo
  {pages} {172} (\bibinfo {year} {2005})},\ \Eprint
  {https://arxiv.org/abs/hep-th/0405069} {arXiv:hep-th/0405069} \BibitemShut
  {NoStop}%
\bibitem [{\citenamefont {Ponomarev}(2023)}]{Ponomarev:2022vjb}%
  \BibitemOpen
  \bibfield  {author} {\bibinfo {author} {\bibfnamefont {D.}~\bibnamefont
  {Ponomarev}},\ }\bibfield  {title} {\bibinfo {title} {{Basic Introduction to
  Higher-Spin Theories}},\ }\href {https://doi.org/10.1007/s10773-023-05399-5}
  {\bibfield  {journal} {\bibinfo  {journal} {Int. J. Theor. Phys.}\ }\textbf
  {\bibinfo {volume} {62}},\ \bibinfo {pages} {146} (\bibinfo {year} {2023})},\
  \Eprint {https://arxiv.org/abs/2206.15385} {arXiv:2206.15385 [hep-th]}
  \BibitemShut {NoStop}%
\bibitem [{\citenamefont {Bekaert}\ \emph {et~al.}(2004)\citenamefont
  {Bekaert}, \citenamefont {Cnockaert}, \citenamefont {Iazeolla},\ and\
  \citenamefont {Vasiliev}}]{Bekaert:2004qos}%
  \BibitemOpen
  \bibfield  {author} {\bibinfo {author} {\bibfnamefont {X.}~\bibnamefont
  {Bekaert}}, \bibinfo {author} {\bibfnamefont {S.}~\bibnamefont {Cnockaert}},
  \bibinfo {author} {\bibfnamefont {C.}~\bibnamefont {Iazeolla}},\ and\
  \bibinfo {author} {\bibfnamefont {M.~A.}\ \bibnamefont {Vasiliev}},\
  }\bibfield  {title} {\bibinfo {title} {{Nonlinear higher spin theories in
  various dimensions}},\ }in\ \href@noop {} {\emph {\bibinfo {booktitle} {{1st
  Solvay Workshop on Higher Spin Gauge Theories}}}}\ (\bibinfo {year} {2004})\
  pp.\ \bibinfo {pages} {132--197},\ \Eprint
  {https://arxiv.org/abs/hep-th/0503128} {arXiv:hep-th/0503128} \BibitemShut
  {NoStop}%
\bibitem [{\citenamefont {Giombi}(2017)}]{Giombi:2016ejx}%
  \BibitemOpen
  \bibfield  {author} {\bibinfo {author} {\bibfnamefont {S.}~\bibnamefont
  {Giombi}},\ }\bibfield  {title} {\bibinfo {title} {{Higher Spin {\textemdash}
  CFT Duality}},\ }in\ \href {https://doi.org/10.1142/9789813149441_0003}
  {\emph {\bibinfo {booktitle} {{Theoretical Advanced Study Institute in
  Elementary Particle Physics}: {New Frontiers in Fields and Strings}}}}\
  (\bibinfo {year} {2017})\ pp.\ \bibinfo {pages} {137--214},\ \Eprint
  {https://arxiv.org/abs/1607.02967} {arXiv:1607.02967 [hep-th]} \BibitemShut
  {NoStop}%
\bibitem [{\citenamefont {Tomasiello}(2024)}]{Tomasiello:2024jyu}%
  \BibitemOpen
  \bibfield  {author} {\bibinfo {author} {\bibfnamefont {A.}~\bibnamefont
  {Tomasiello}},\ }\bibfield  {title} {\bibinfo {title} {{Higher spins and
  Finsler geometry}},\ }\href {https://doi.org/10.1007/JHEP10(2024)047}
  {\bibfield  {journal} {\bibinfo  {journal} {JHEP}\ }\textbf {\bibinfo
  {volume} {10}},\ \bibinfo {pages} {047}},\ \Eprint
  {https://arxiv.org/abs/2405.00776} {arXiv:2405.00776 [hep-th]} \BibitemShut
  {NoStop}%
\bibitem [{\citenamefont {Addazi}\ \emph {et~al.}(2022)\citenamefont {Addazi}
  \emph {et~al.}}]{Addazi:2021xuf}%
  \BibitemOpen
  \bibfield  {author} {\bibinfo {author} {\bibfnamefont {A.}~\bibnamefont
  {Addazi}} \emph {et~al.},\ }\bibfield  {title} {\bibinfo {title} {{Quantum
  gravity phenomenology at the dawn of the multi-messenger era\textemdash{}A
  review}},\ }\href {https://doi.org/10.1016/j.ppnp.2022.103948} {\bibfield
  {journal} {\bibinfo  {journal} {Prog. Part. Nucl. Phys.}\ }\textbf {\bibinfo
  {volume} {125}},\ \bibinfo {pages} {103948} (\bibinfo {year} {2022})},\
  \Eprint {https://arxiv.org/abs/2111.05659} {arXiv:2111.05659 [hep-ph]}
  \BibitemShut {NoStop}%
\bibitem [{\citenamefont {Colladay}\ and\ \citenamefont
  {Kostelecky}(1998)}]{Colladay:1998fq}%
  \BibitemOpen
  \bibfield  {author} {\bibinfo {author} {\bibfnamefont {D.}~\bibnamefont
  {Colladay}}\ and\ \bibinfo {author} {\bibfnamefont {V.~A.}\ \bibnamefont
  {Kostelecky}},\ }\bibfield  {title} {\bibinfo {title} {{Lorentz violating
  extension of the standard model}},\ }\href
  {https://doi.org/10.1103/PhysRevD.58.116002} {\bibfield  {journal} {\bibinfo
  {journal} {Phys. Rev.}\ }\textbf {\bibinfo {volume} {D58}},\ \bibinfo {pages}
  {116002} (\bibinfo {year} {1998})},\ \Eprint
  {https://arxiv.org/abs/hep-ph/9809521} {arXiv:hep-ph/9809521 [hep-ph]}
  \BibitemShut {NoStop}%
\bibitem [{\citenamefont {Kostelecky}(2011)}]{Kostelecky:2011qz}%
  \BibitemOpen
  \bibfield  {author} {\bibinfo {author} {\bibfnamefont {A.}~\bibnamefont
  {Kostelecky}},\ }\bibfield  {title} {\bibinfo {title} {{Riemann-Finsler
  geometry and Lorentz-violating kinematics}},\ }\href
  {https://doi.org/10.1016/j.physletb.2011.05.041} {\bibfield  {journal}
  {\bibinfo  {journal} {Phys. Lett.}\ }\textbf {\bibinfo {volume} {B701}},\
  \bibinfo {pages} {137} (\bibinfo {year} {2011})},\ \Eprint
  {https://arxiv.org/abs/1104.5488} {arXiv:1104.5488 [hep-th]} \BibitemShut
  {NoStop}%
\bibitem [{\citenamefont {Amelino-Camelia}(2002)}]{Amelino-Camelia2002b}%
  \BibitemOpen
  \bibfield  {author} {\bibinfo {author} {\bibfnamefont {G.}~\bibnamefont
  {Amelino-Camelia}},\ }\bibfield  {title} {\bibinfo {title} {{Doubly special
  relativity}},\ }\href {https://doi.org/10.1038/418034a} {\bibfield  {journal}
  {\bibinfo  {journal} {Nature}\ }\textbf {\bibinfo {volume} {418}},\ \bibinfo
  {pages} {34} (\bibinfo {year} {2002})},\ \Eprint
  {https://arxiv.org/abs/gr-qc/0207049} {arXiv:gr-qc/0207049 [gr-qc]}
  \BibitemShut {NoStop}%
\bibitem [{\citenamefont {Amelino-Camelia}(2001)}]{AmelinoCamelia:2001vy}%
  \BibitemOpen
  \bibfield  {author} {\bibinfo {author} {\bibfnamefont {G.}~\bibnamefont
  {Amelino-Camelia}},\ }\bibfield  {title} {\bibinfo {title} {{Status of
  relativity with observer independent length and velocity scales}},\ }\href
  {https://doi.org/10.1063/1.1419321} {\bibfield  {journal} {\bibinfo
  {journal} {AIP Conf.Proc.}\ }\textbf {\bibinfo {volume} {589}},\ \bibinfo
  {pages} {137} (\bibinfo {year} {2001})},\ \Eprint
  {https://arxiv.org/abs/gr-qc/0106004} {arXiv:gr-qc/0106004 [gr-qc]}
  \BibitemShut {NoStop}%
\bibitem [{\citenamefont {Amelino-Camelia}(2013)}]{AmelinoCamelia:2008qg}%
  \BibitemOpen
  \bibfield  {author} {\bibinfo {author} {\bibfnamefont {G.}~\bibnamefont
  {Amelino-Camelia}},\ }\bibfield  {title} {\bibinfo {title}
  {{Quantum-Spacetime Phenomenology}},\ }\href
  {https://doi.org/10.12942/lrr-2013-5} {\bibfield  {journal} {\bibinfo
  {journal} {Living Rev.Rel.}\ }\textbf {\bibinfo {volume} {16}},\ \bibinfo
  {pages} {5} (\bibinfo {year} {2013})},\ \Eprint
  {https://arxiv.org/abs/0806.0339} {arXiv:0806.0339 [gr-qc]} \BibitemShut
  {NoStop}%
\bibitem [{\citenamefont {Lammerzahl}\ \emph {et~al.}(2005)\citenamefont
  {Lammerzahl}, \citenamefont {Macias},\ and\ \citenamefont
  {Mueller}}]{Lammerzahl:2005jw}%
  \BibitemOpen
  \bibfield  {author} {\bibinfo {author} {\bibfnamefont {C.}~\bibnamefont
  {Lammerzahl}}, \bibinfo {author} {\bibfnamefont {A.}~\bibnamefont {Macias}},\
  and\ \bibinfo {author} {\bibfnamefont {H.}~\bibnamefont {Mueller}},\
  }\bibfield  {title} {\bibinfo {title} {{Lorentz invariance violation and
  charge (non-)conservation: A General theoretical frame for extensions of the
  Maxwell equations}},\ }\href {https://doi.org/10.1103/PhysRevD.71.025007}
  {\bibfield  {journal} {\bibinfo  {journal} {Phys. Rev. D}\ }\textbf {\bibinfo
  {volume} {71}},\ \bibinfo {pages} {025007} (\bibinfo {year} {2005})},\
  \Eprint {https://arxiv.org/abs/gr-qc/0501048} {arXiv:gr-qc/0501048}
  \BibitemShut {NoStop}%
\bibitem [{\citenamefont {Xiao}\ \emph {et~al.}(2010)\citenamefont {Xiao},
  \citenamefont {Shao},\ and\ \citenamefont {Ma}}]{Xiao:2010yx}%
  \BibitemOpen
  \bibfield  {author} {\bibinfo {author} {\bibfnamefont {Z.}~\bibnamefont
  {Xiao}}, \bibinfo {author} {\bibfnamefont {L.}~\bibnamefont {Shao}},\ and\
  \bibinfo {author} {\bibfnamefont {B.-Q.}\ \bibnamefont {Ma}},\ }\bibfield
  {title} {\bibinfo {title} {{Eikonal equation of the Lorentz-violating Maxwell
  theory}},\ }\href {https://doi.org/10.1140/epjc/s10052-010-1502-4} {\bibfield
   {journal} {\bibinfo  {journal} {Eur. Phys. J. C}\ }\textbf {\bibinfo
  {volume} {70}},\ \bibinfo {pages} {1153} (\bibinfo {year} {2010})},\ \Eprint
  {https://arxiv.org/abs/1011.5074} {arXiv:1011.5074 [hep-th]} \BibitemShut
  {NoStop}%
\bibitem [{\citenamefont {Casana}\ \emph {et~al.}(2011)\citenamefont {Casana},
  \citenamefont {Ferreira},\ and\ \citenamefont {Moreira}}]{Casana:2011du}%
  \BibitemOpen
  \bibfield  {author} {\bibinfo {author} {\bibfnamefont {R.}~\bibnamefont
  {Casana}}, \bibinfo {author} {\bibfnamefont {M.~M.}\ \bibnamefont {Ferreira},
  \bibfnamefont {Jr}},\ and\ \bibinfo {author} {\bibfnamefont {R.~P.~M.}\
  \bibnamefont {Moreira}},\ }\bibfield  {title} {\bibinfo {title} {{Aspects of
  a planar nonbirefringent and CPT-even electrodynamics stemming from the
  Standard Model Extension}},\ }\href
  {https://doi.org/10.1103/PhysRevD.84.125014} {\bibfield  {journal} {\bibinfo
  {journal} {Phys. Rev. D}\ }\textbf {\bibinfo {volume} {84}},\ \bibinfo
  {pages} {125014} (\bibinfo {year} {2011})},\ \Eprint
  {https://arxiv.org/abs/1108.6193} {arXiv:1108.6193 [hep-th]} \BibitemShut
  {NoStop}%
\bibitem [{\citenamefont {Chang}\ and\ \citenamefont
  {Wang}(2012)}]{Chang:2012ks}%
  \BibitemOpen
  \bibfield  {author} {\bibinfo {author} {\bibfnamefont {Z.}~\bibnamefont
  {Chang}}\ and\ \bibinfo {author} {\bibfnamefont {S.}~\bibnamefont {Wang}},\
  }\bibfield  {title} {\bibinfo {title} {{Lorentz invariance violation and
  electromagnetic field in an intrinsically anisotropic spacetime}},\ }\href
  {https://doi.org/10.1140/epjc/s10052-012-2165-0} {\bibfield  {journal}
  {\bibinfo  {journal} {Eur. Phys. J. C}\ }\textbf {\bibinfo {volume} {72}},\
  \bibinfo {pages} {2165} (\bibinfo {year} {2012})},\ \Eprint
  {https://arxiv.org/abs/1204.2478} {arXiv:1204.2478 [hep-ph]} \BibitemShut
  {NoStop}%
\bibitem [{\citenamefont {Cri{\c{s}}an}\ and\ \citenamefont
  {Vancea}(2020)}]{Crisan:2020krc}%
  \BibitemOpen
  \bibfield  {author} {\bibinfo {author} {\bibfnamefont {A.~V.}\ \bibnamefont
  {Cri{\c{s}}an}}\ and\ \bibinfo {author} {\bibfnamefont {I.~V.}\ \bibnamefont
  {Vancea}},\ }\bibfield  {title} {\bibinfo {title} {{Finsler geometries from
  topological electromagnetism}},\ }\href
  {https://doi.org/10.1140/epjc/s10052-020-8123-3} {\bibfield  {journal}
  {\bibinfo  {journal} {Eur. Phys. J. C}\ }\textbf {\bibinfo {volume} {80}},\
  \bibinfo {pages} {566} (\bibinfo {year} {2020})},\ \Eprint
  {https://arxiv.org/abs/2006.03865} {arXiv:2006.03865 [hep-th]} \BibitemShut
  {NoStop}%
\bibitem [{\citenamefont {Takka}\ and\ \citenamefont
  {Bouda}(2019)}]{Takka:2019him}%
  \BibitemOpen
  \bibfield  {author} {\bibinfo {author} {\bibfnamefont {N.}~\bibnamefont
  {Takka}}\ and\ \bibinfo {author} {\bibfnamefont {A.}~\bibnamefont {Bouda}},\
  }\bibfield  {title} {\bibinfo {title} {{Maxwell{\textquoteright}s equations
  and Lorentz force in doubly special relativity}},\ }\href
  {https://doi.org/10.1007/s12648-019-01556-x} {\bibfield  {journal} {\bibinfo
  {journal} {Indian J. Phys.}\ }\textbf {\bibinfo {volume} {94}},\ \bibinfo
  {pages} {1227} (\bibinfo {year} {2019})},\ \Eprint
  {https://arxiv.org/abs/2207.14531} {arXiv:2207.14531 [gr-qc]} \BibitemShut
  {NoStop}%
\bibitem [{\citenamefont {Relancio}\ and\ \citenamefont
  {Santamar{\'\i}a-Sanz}(2025)}]{Relancio:2024nud}%
  \BibitemOpen
  \bibfield  {author} {\bibinfo {author} {\bibfnamefont {J.~J.}\ \bibnamefont
  {Relancio}}\ and\ \bibinfo {author} {\bibfnamefont {L.}~\bibnamefont
  {Santamar{\'\i}a-Sanz}},\ }\bibfield  {title} {\bibinfo {title} {{Generalized
  Hamilton spaces: New developments and applications}},\ }\href
  {https://doi.org/10.1016/j.geomphys.2025.105626} {\bibfield  {journal}
  {\bibinfo  {journal} {J. Geom. Phys.}\ }\textbf {\bibinfo {volume} {217}},\
  \bibinfo {pages} {105626} (\bibinfo {year} {2025})},\ \Eprint
  {https://arxiv.org/abs/2407.18819} {arXiv:2407.18819 [math-ph]} \BibitemShut
  {NoStop}%
\end{thebibliography}
\end{document}